\documentclass{ws-procs9x6}

\begin{document}

\title{Search for Pentaquark States with CLAS at Jefferson Lab}

\author{V. D. Burkert}
\address{Jefferson Lab, 12000 Jefferson Avenue, Newport News VA 23606, USA\\
 E-mail: burkert@jlab.org}

\author{R. De Vita}
\address{Istituto Nazionale di Fisica Nucleare\\
via Dodecaneso 33, 16146 Genova, Italy\\ 
E-mail: devita@ge.infn.it}

\author{S. Niccolai}
\address{IPN Orsay, \\
15 rue Georges Clemenceau,\\
91406 Orsay France\\
E-mail: silvia@jlab.org}

\author{and the CLAS Collaboration}

\maketitle

\abstracts{We discuss the experimental program to search for baryon 
states with exotic flavor quantum numbers using CLAS at Jefferson Lab}

\section{Physics Motivation}

\noindent
The existence of baryons with quantum numbers that cannot be obtained with 
only 3 valence quarks $qqq$, but require a minimum quark content $qqqq\bar{q}$, 
has excited the hadron community since the first public announcements of such a 
state were made in the year 2003. The observed state is now called the $\Theta^+(1540)$.
It appears to have a mass in the range 1525 MeV to 1550 MeV, and strangeness S=+1~\cite{overview}. 
While pentaquark states 
with such quantum numbers have been discussed for years, specific predictions for both
a mass of 1530 MeV and a narrow width of $<$ 15 MeV were made in a paper 
by Diakonov et al. in 1997~\cite{diakonov97}, 
based on the Chiral Soliton Model ($\chi SM$). In this model, the $\Theta^+$ is an 
isosinglet member of a $J={1\over 2}^+$ anti-decuplet of ten states with a minimum 
quark content of 5 quarks (``pentaquarks''), three of the states have exotic flavor 
quantum numbers, the $\Theta^+$, and the two baryons $\Xi_5^{- -}$ and $\Xi_5^+$. 
The other non-exotic members are three $\Sigma_5$, and two $N_5^*$ states. 
The width appears to be much more narrow than the experimental resolutions, and 
may be as narrow as 1 MeV\cite{arndt,cahn,gibbs}.
Following the first experimental announcements, an avalanche of theoretical papers have appeared 
trying to understand its low mass and narrow width, as well as to make predictions on
production mechanism and possible excited states of the $\Theta^+$~\cite{stancu,oka}. 
Lattice QCD is currently not providing fully satisfactory predictions for the $\Theta^+$.  
One group finds no signal, three groups find a signal at about the right mass, two at 
negative parity, one at positive parity\cite{kovacs}.

Evidence for the state has been claimed in more than 10 published 
works~\cite{hicks} from medium energy to very high energy experiments. 
There are a number of experiments, mostly at high energies, that report null results. 
Most of the results, if not all, come from the analysis of data that were taken for 
other purposes. This fact may explain the relatively low significance of all positive 
results, which ranges from about 4$\sigma$ to 7$\sigma$ for individual experiments.  
Experimentally, the $\Theta^+$ has been observed in either $nK^+$, or $pK^{0}$ final 
states. Reported masses in some cases vary by more than the uncertainties given for the individual 
experiments, with the masses obtained from processes involving 
$nK^+$ in the initial or final states giving on average 10-15 MeV higher masses. 
The discrepancy in mass determination needs 
to be resolved, but may not solely be an experimental problem. For example, the
 mass difference might be explained by different initial or final state interactions involved 
in $nK^+$ and $pK^{0}$ channels. Also, different interference effects could be involved that depend on the 
kinematics where the signal is observed. In either case theoretical input will be needed to resolve
the discrepancy.  
Finding a definite answer to the question of existence or non-existence of the $\Theta^+$ and of the other
 5-quark baryons is, of course, of overriding importance and urgency. It will tell us much about 
how QCD works at the hadron scale, and can only be answered experimentally. 

In this 
contribution we report on the results from Jefferson Lab using the CLAS detector, 
and on the current program of second generation experiments aimed at improving the 
statistical accuracy of the measurements by at least one order of magnitude. We also need to 
better understand the systematics involved and obtain some insight into production mechanisms. 
In  section~\ref{sect:clas} we report the experimental apparatus to the degree it is relevant 
for the physics at hand. In section~\ref{clas_theta} we discuss the already published 
results. Section~\ref{ongoing_analysis} describes ongoing analyses, and in section~\ref{perspectives} 
we report on the status of two second generation measurements, and then outline planned experiments 
to further study the systematics of pentaquark states.

\begin{figure}
\vspace{6.cm}
\includegraphics{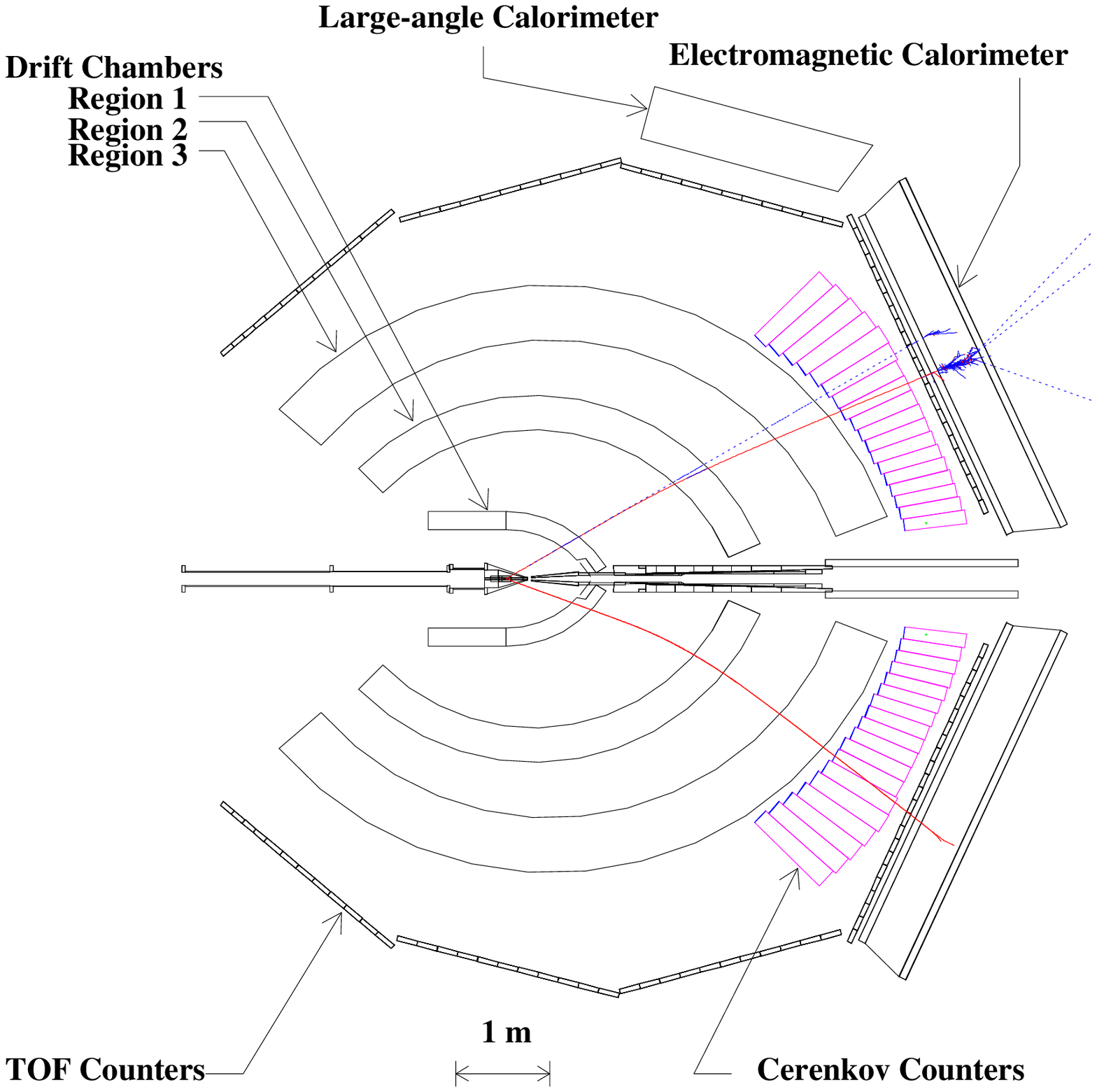}
\includegraphics{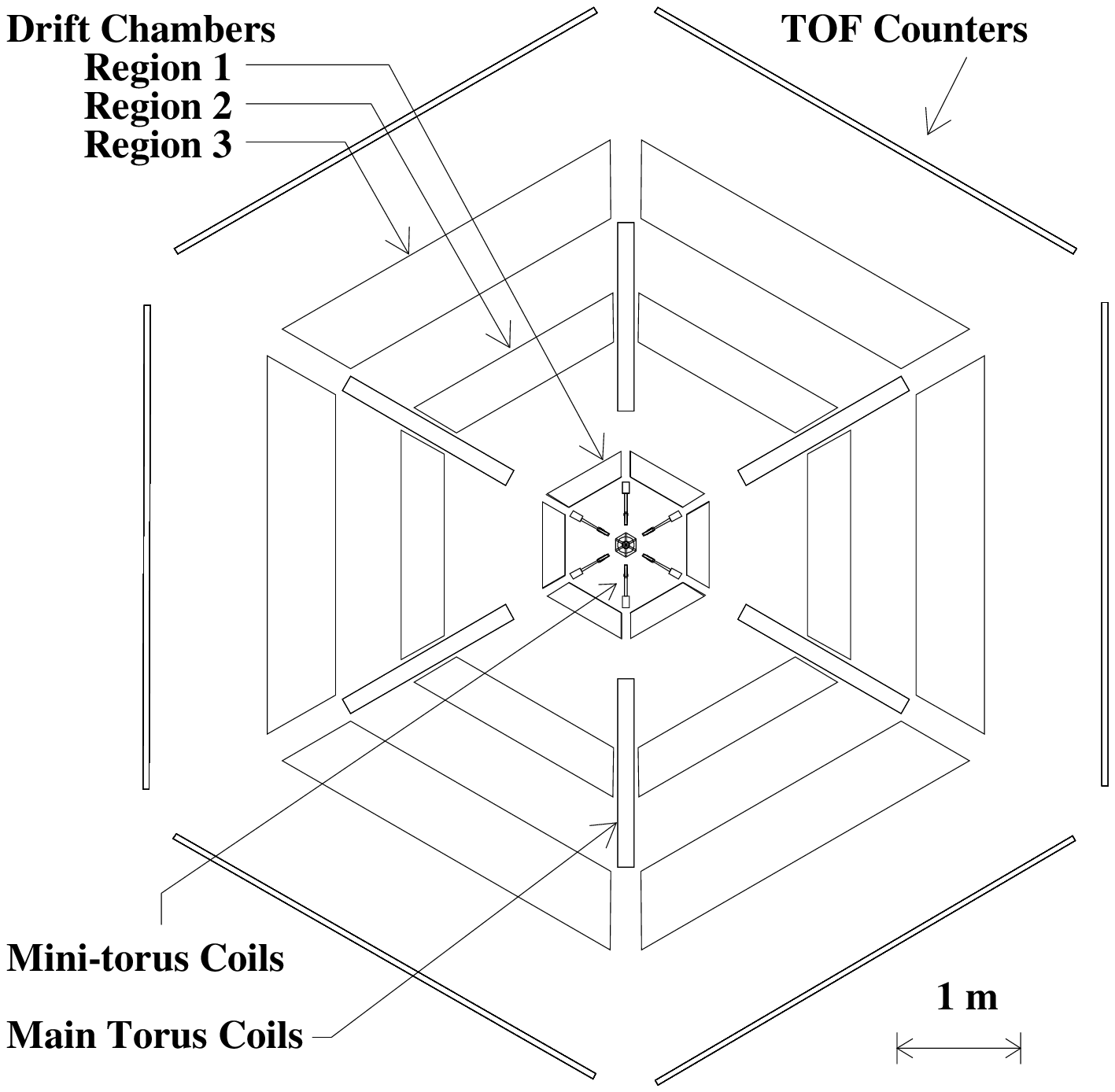}
\caption{The CLAS detector. Left: Longitudinal cut along the beam line shows the 3 drift 
chamber regions, the Cherenkov counters at forward angles for electron and pion separation, 
the time-of-flight system, and the electromagnetic 
calorimeters for the detection of photons and neutrons. Right: Transverse cut through CLAS. 
The superconducting torus coils provide a six 
sector structure, each sector being instrumented with independent detectors.} 
\label{fig:clas}
\end{figure}

\section{CLAS and CEBAF - unique capabilities for baryon spectroscopy \label{sect:clas}}

The continuous electron beam provided by the CEBAF accelerator is
converted into a bremsstrahlung photon beam at CLAS using a gold radiator located 
20 meters upstream of a liquid hydrogen or deuterium target. The photon energy 
is measured by detecting the scattered the electrons that generated the photons in 
the energy range from 20\% to 95\% of the incident electron energy. 
Typical photon beam rates range from a few times $10^{6}$ to  $10^7$ per
second. The  CLAS detector is shown in Fig.~\ref{fig:clas}. At its core is a superconducting
toroidal magnet, providing momentum analysis of charged particles in six 
sectors.  The magnetic field produced by the six-coil toroidal magnet 
is oriented in such a way as to maintain a constant azimuthal angle of the 
scattered particles while changing  only their polar angle.
Tracking is provided with three regions of drift chambers with a total of 34 layers 
of drift cells arranged radially from the target. The total number of drift cells 
in CLAS is about 35,000 providing a highly redundant tracking information. 
Charged tracks are reconstructed in polar angles from about 10 to 140 degrees. 
288 plastic scintillator paddles 
provide time-of-flight information used for particle identification. 
For experiments with a bremsstrahlung photon beam, a segmented scintillation counter is 
arranged around the target for triggering and to provide improved start time information. 
CLAS is optimized for the selection of exclusive processes in a large kinematic range 
and with good resolution. The large 
acceptance of CLAS allows simultaneous measurement of  several processes. 

Often, as is also the case in the study of the $\Theta^+$,
 the missing mass technique is used to identify final states with one unmeasured 
particle, usually a neutron or a neutral meson. At intermediate energies all 
kinematic regions can play important roles, and may be sensitive to 
different production mechanisms, e.g. in s-channel, t-channel, and u-channel 
processes. 
With its large coverage CLAS can explore baryon excitation processes in 
all of these regions. These capabilities are unique to CLAS and, as long as the 
major production mechanisms are unknown, are crucial in the 
study of processes with small cross sections, such as the $\Theta^+$ or its possible excited 
states. For example, the small $\Theta^+$ signal may be completely swamped by 
background processes in t-channel kinematics. However, it may show up more prominently 
in u-channel kinematics through baryon exchange processes at large cms angles. 
The wide coverage of CLAS allows selection of all kinematics and reaction channels, 
and provides the utmost in sensitivity.  

\begin{table}
\tbl{Runs completed in Hall B at Jefferson Lab with the CLAS detector using photon beams. 
The experimental conditions are summarized.}
{ \begin{tabular}{|c|c|c|c|c|}
\hline
{Experiment} & {Year} & {Beam Energy} & {Target} & {$\int{\mathcal{L}}$ (pb$^{-1}$)} \\ 
\hline\hline
{g6a} & {1998} & {4.1 GeV} & {LH$_2$} & {$\sim$ 1.0}  \\
\hline
{g6b} & {1999} & {5.5 GeV} & {LH$_2$} & {$\sim$ 1.0}   \\
\hline
{g1c} & {1999} & {1.9-3.1 GeV} & {LH$_2$} & {$\sim$ 4.0}  \\
\hline
{g6c} & {2000} & {5.7 GeV} & {LH$_2$} & {$\sim$ 2.7}  \\
\hline
{g2a} & {1999} & {2.5-3.1 GeV} & {LD$_2$} & {$\sim$ 2}  \\
\hline
{g3a} & {1999} & {1.6 GeV} & {L$^3$He} & {$\sim$ 0.3}   \\
\hline
\end{tabular}
\label{table:runs}}
\end{table}

Since 1998, several experiments using photon beams and different 
targets have been completed: a summary is given in Table \ref{table:runs}.
In the last year, the existing CLAS data were reanalyzed to study 
possible evidence for pentaquark production. 
The CLAS collaboration has published two positive signals on the $\Theta^+$ 
using deuterium~\cite{clas_d} and hydrogen targets~\cite{clas_p}.  
We discuss these in the following section.

\section{Discussion of published results \label{clas_theta}}

\subsection{Production on deuterium}

\begin{figure}[hbpt]
\vspace{5.cm}
\includegraphics{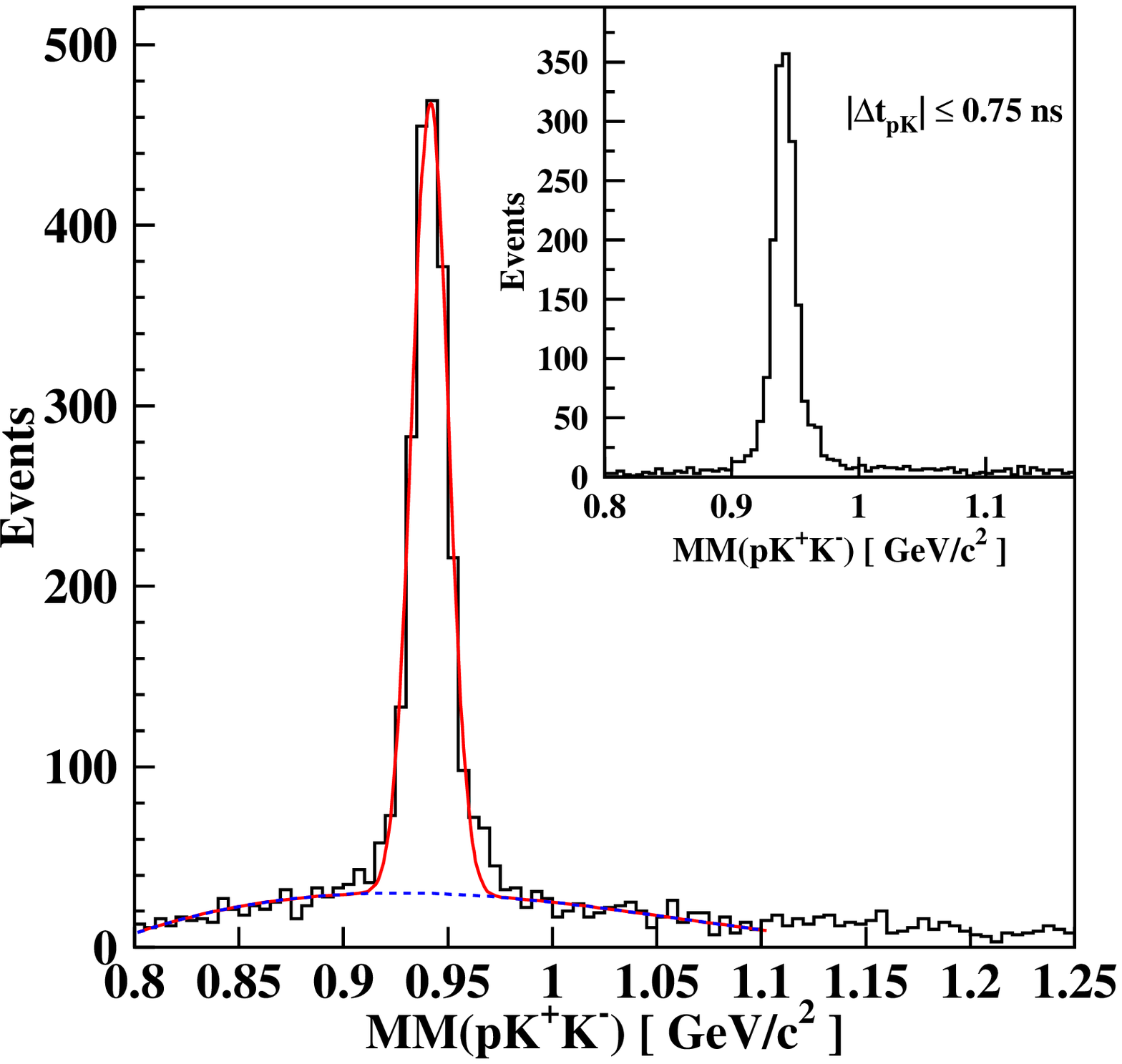}
\includegraphics{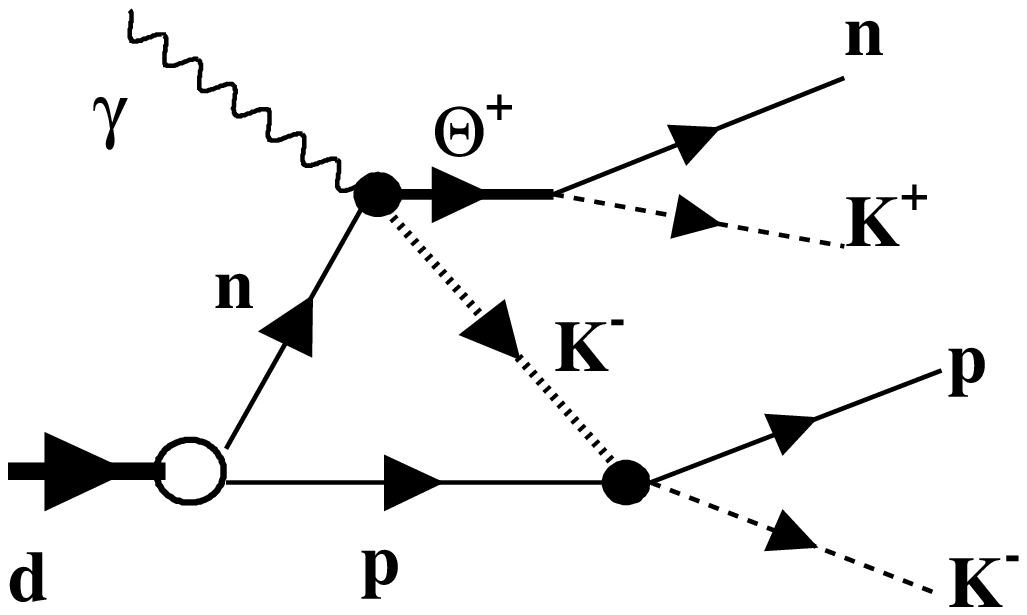}
\caption{Left: Missing mass $M_X$ of $\gamma d \rightarrow pK^-K^+ X$. A peak at the
neutron mass is seen. The inset shows the results with more stringent vertex time cuts.
Right: Possible diagram for the observed process.}  
\label{fig:clas_d_diagram}
\label{fig:clas_d_neutron}
\end{figure}

\begin{figure}[hbpt]
\vspace{5.cm}
\includegraphics{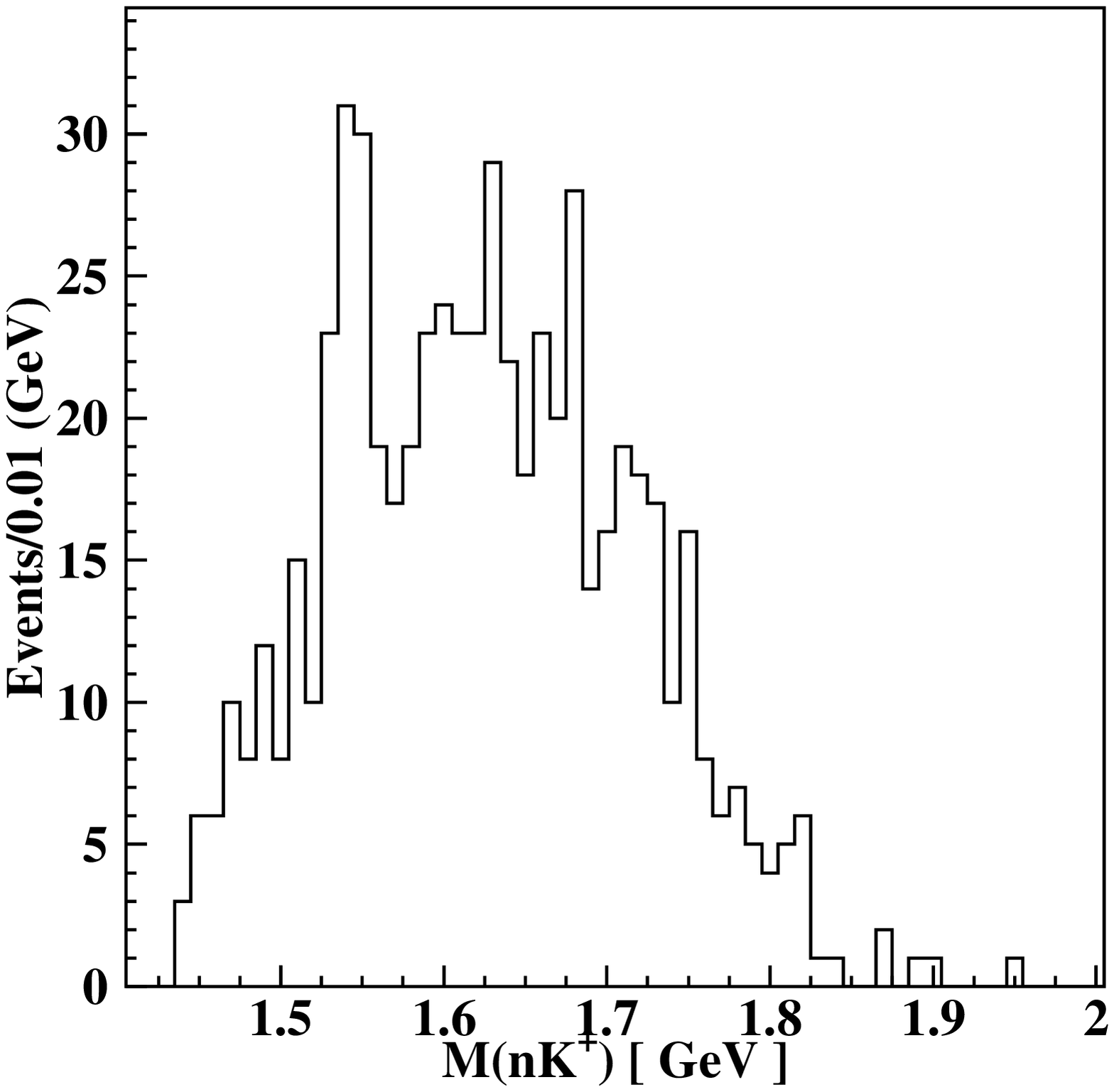}
\includegraphics{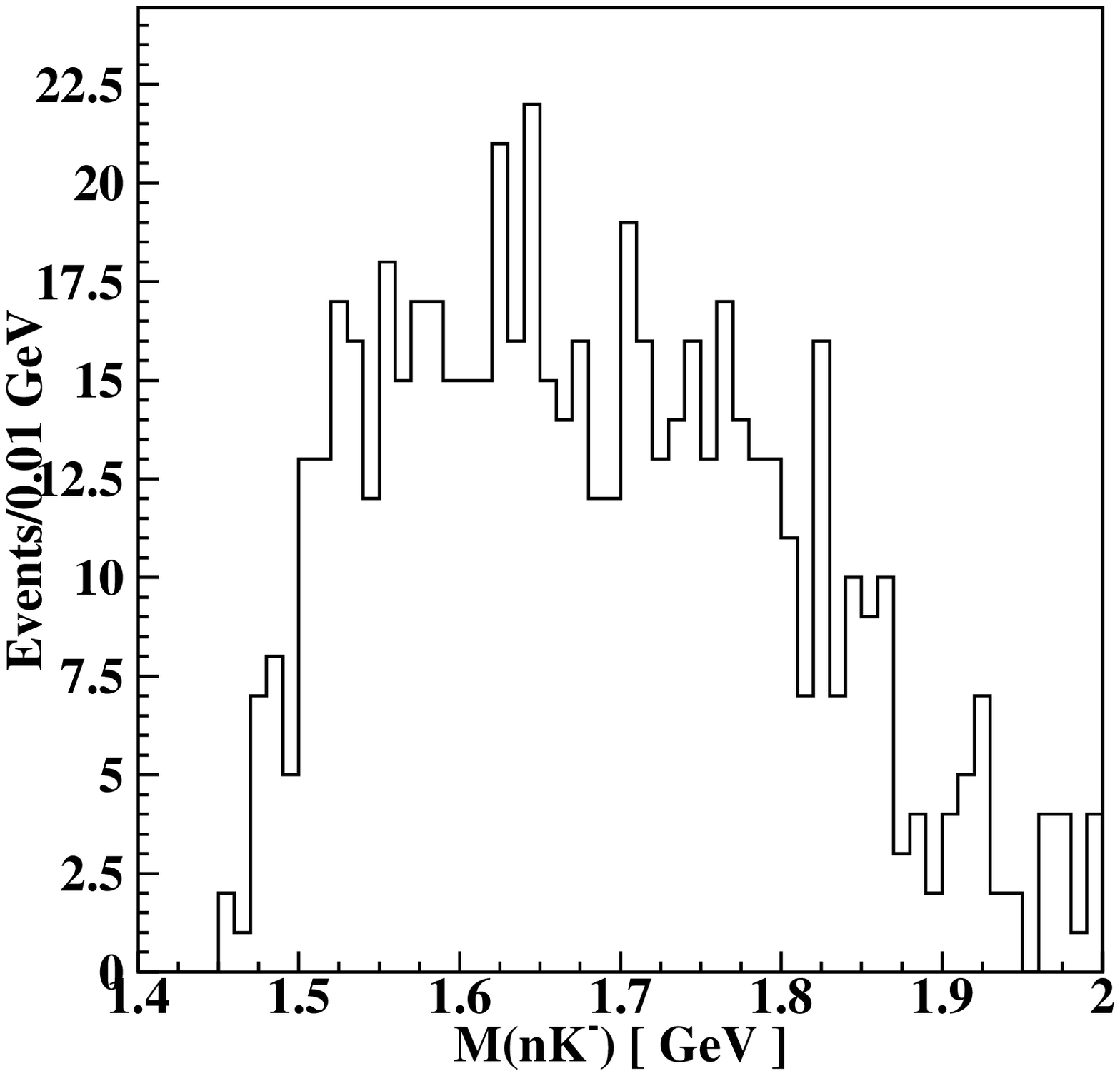}
\caption{Left: Invariant mass distribution $M(nK^+)$ for the final state $pK^-nK^+$.
Right:$M(nK^-)$ for the same event as in the left graph. The lack of a narrow peak shows that 
kinematic reflections do not seem to generate narrow structures in $nK^-$.} 
\label{fig:clas_d_nk+}
\label{fig:clas_d_nk-}
\end{figure}

The first evidence for the $\Theta^+$ was seen in experiments on nuclear 
targets~\cite{leps,diana,clas_d}.  In CLAS, the fully exclusive process 
$\gamma D \rightarrow K^-pK^+n$ was measured. The 4-momentum vectors 
of the photon, target nucleus, and all charged particles in the final state are 
known, the neutron can be identified by computing the missing mass of the 
remaining system as can be seen in Fig.~\ref{fig:clas_d_neutron}. 
In this process the $\Theta^+$ would be produced on the neutron in the 
deuteron, while the proton would be a spectator. However, in order to be 
able to detect the proton in CLAS, it must have a minimum momentum 
of about 200 MeV/c. This requires a complex final state interaction. The final 
state also contains $\phi$ and $\Lambda(1520)$ that are eliminated by cuts 
in the $K^+K^-$ and $pK^-$ invariant mass spectra. A possible diagram is shown 
in Fig.~\ref{fig:clas_d_diagram}. The final spectrum is shown in 
Figure~\ref{fig:clas_d_nk+}. The $nK^+$ invariant mass shows a significant peak 
at $1542 \pm 5$~MeV. An analysis of these data using a different technique finds 
that the significance of the observed peak may not be as large as presented in the published
work. We expect a
 definitive answer from a much larger statistics data set (g10) that
 is currently being analyzed.

. 

Could this peak be generated by a statistical fluctuation? If we consider 
this result isolated from the evidence seen by other independent
 experiments, there is a small probability that it could be a 
fluctuation. However, such an interpretation becomes extremely unlikely if 
the evidence from other independent experiments is considered. 
Other effects such as kinematical reflections have been studied. 
Here structures in (NK) mass spectra may be produced as kinematical reflections 
of high mass meson production with decays to the $K^+K^-$ final state. 
If  the phase space for the reaction is limited, the invariant mass $nK^+$ could 
show an enhancement in the 1.55 GeV mass range. Kinematical reflections 
are well-known effects in spectroscopy, and have been studied extensively. 
For the current analysis, broad enhancements or shoulders can indeed be 
produced this way as is shown in the right panel of Fig.~\ref{fig:clas_d_nk-}, 
where the mass of the $K^-n$ is shown using the same events that 
are included in the left panel. The smooth shoulder near 1.55 GeV may indeed be
due to such an effect. Clearly, the sharp $\Theta^+$ peak in the left panel has very
different characteristics from a kinematical reflection.        

\subsection{Production on hydrogen}

\begin{figure}
\vspace{5.cm}
\includegraphics{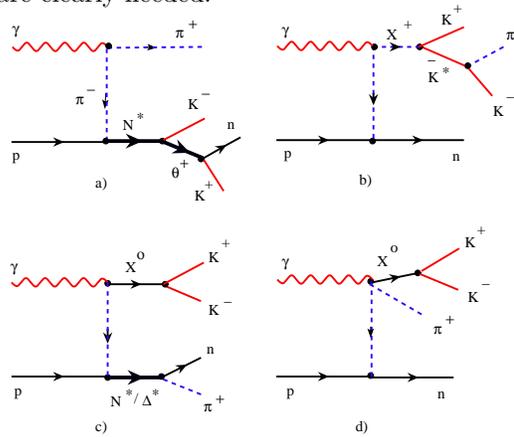}
\caption{ Diagrams that may contribute to the process $\gamma p \rightarrow \pi^+K^-nK^+$.
The left top diagram contributes to $\Theta^+$ production through intermediate $N^*$ excitation. The other
diagrams represent background processes.} 
\label{fig:diagram_p}
\end{figure}

\begin{figure}
\vspace{5.cm}
\includegraphics{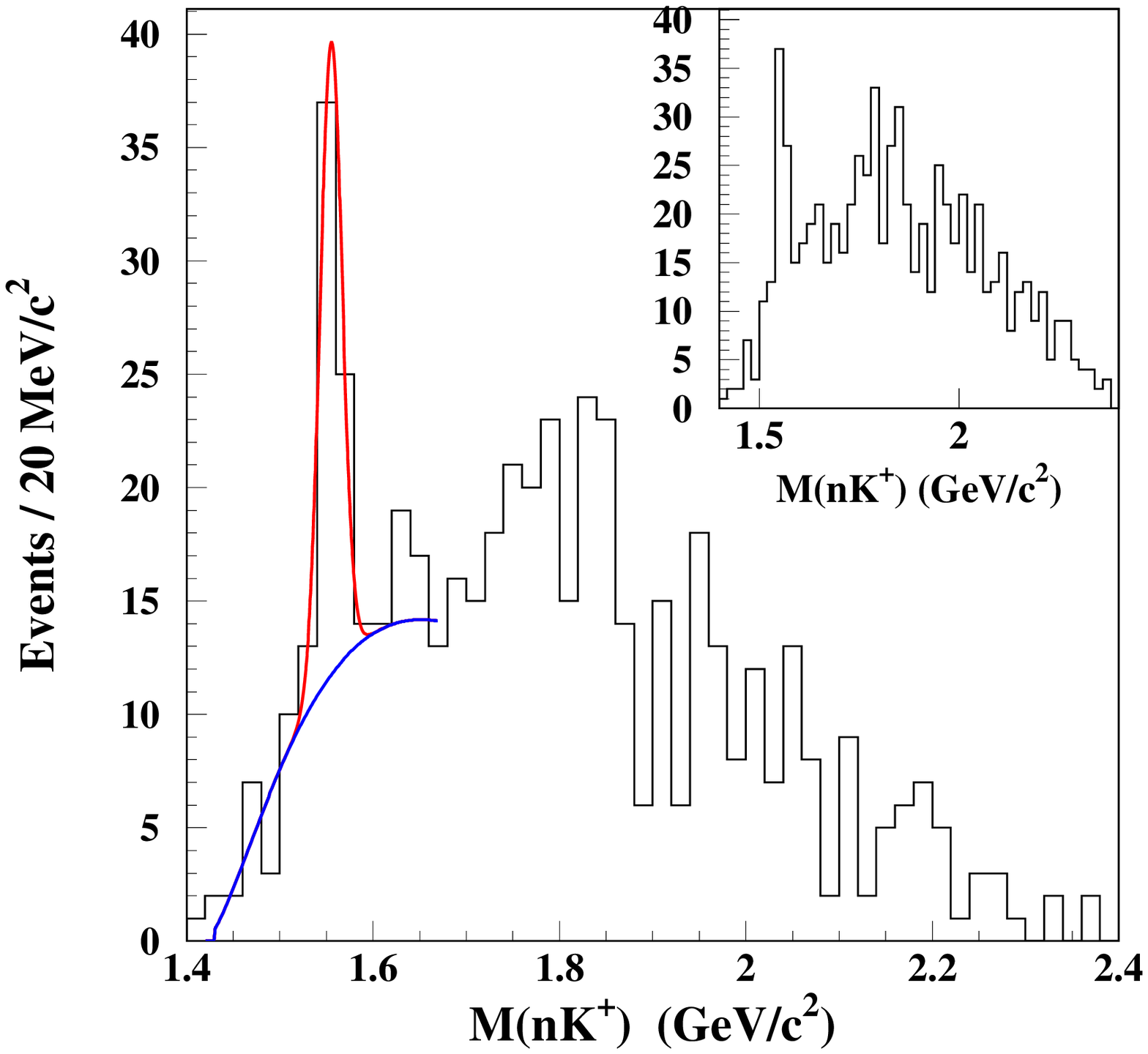}
\includegraphics{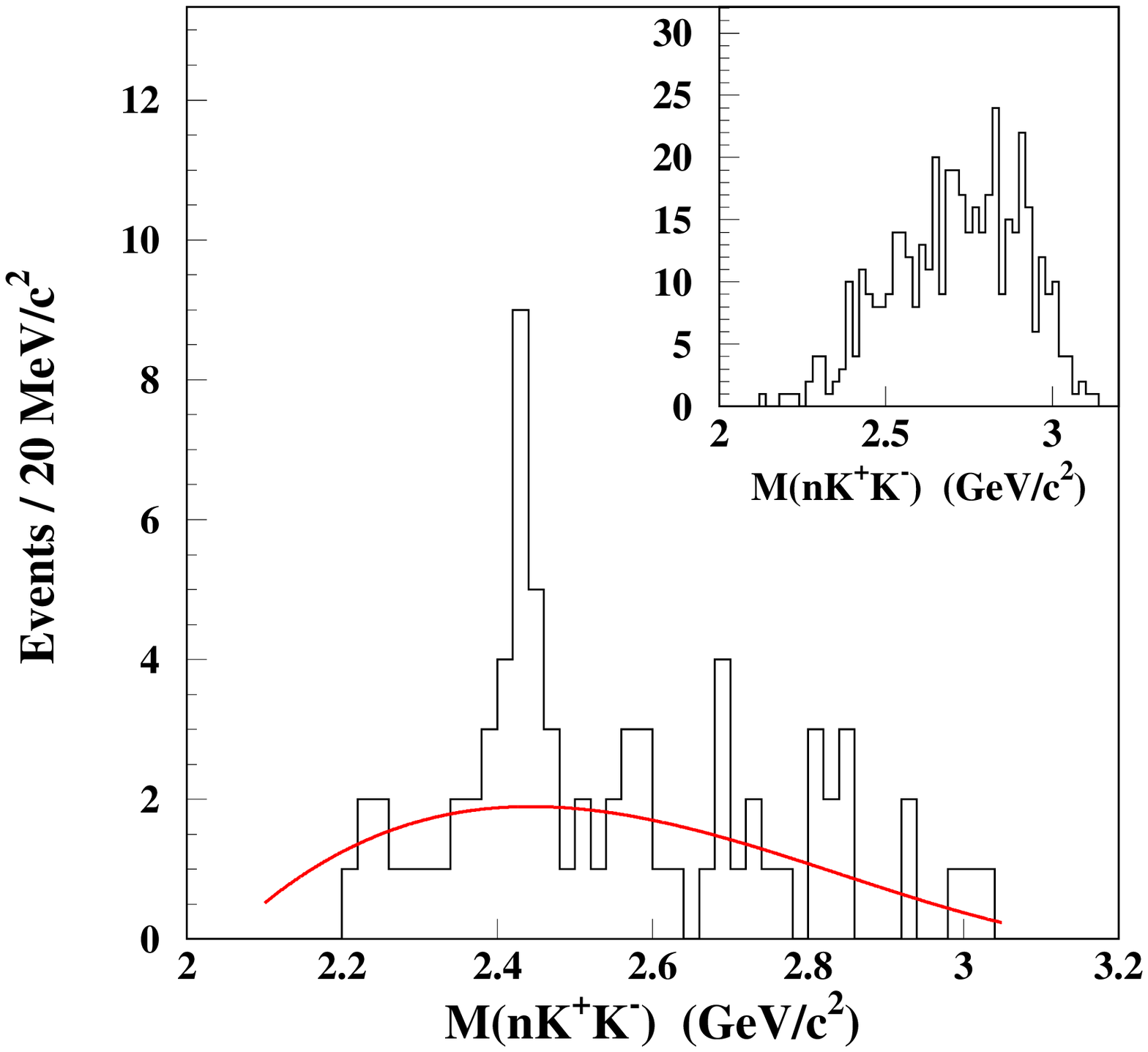}
\caption{Left: Invariant mass distribution of $M(nK^+)$ after all cuts. The inset shows the $nK^+$ 
mass distribution
with only the $\cos\theta^*_{\pi^+} > 0.8$ cut applied. Right: Mass distribution $M(K^-nK^+)$ for events selected in the peak region of the graph on the left. The inset shows the distribution for events outside of the $\Theta^+$ region.} 
\label{fig:clas_p}
\end{figure}

 Here the process $\gamma p \rightarrow \pi^+ K^+K^- n$ is 
selected, using the g6a, g6b, and g6c data runs~\cite{clas_p}. 
Cuts are applied with the hypothesis that the $\Theta^+$ is produced 
via intermediate $N^*$ excitation. 
Possible contributions to that channel are shown in Fig.~\ref{fig:diagram_p}. 
Events are selected with a forward angle $\pi^+$, and events with a forward 
angle $K^+$ are ejected. The latter cut was used to reduce t-channel 
contributions to $K^+$. The final mass spectrum is shown in Fig.~\ref{fig:clas_p}. 
A very significant peak is seen at a mass of $1.555 \pm 10$ MeV. In this 
analysis background processes contributing to the selected channel were subjected to 
a partial wave analysis. This allowed for a precise determination of the shape of
the background contributions. 

If the reaction mechanism is indeed through excitation of 
an intermediate $N^*$, then such a $N^*$ would presumably not be a usual
 3-quark state but rather a 5-quark or even 7-quark state with strong
 $s\bar{s}$ components.  Selecting events in the mass range around the 
peak, and plotting the invariant mass $nK^+K^-$ in Fig.~\ref{fig:clas_p} 
we see indications of a structure near 2.4 GeV. While the results hint 
at a narrow structure near 2.4 GeV the statistics are too poor to allow 
drawing definite conclusions.
 Further studies with higher statistics are clearly needed.  

\section{Ongoing analysis \label{ongoing_analysis}}

{\sl In this section we describe analyses that have shown some promise in the search for the $\Theta^+$ and possible
excited states, however they lack the statistics and significance to be fully convincing. 
The CLAS collaboration has decided to not publish these works but wait for the completion of the ongoing 
runs with much higher statistics. Although no final results are presented here, it may be instructive to describe 
the techniques used in preparation for the analysis of the new high statistics data. }    

\subsection{Reactions on protons}

The $g1c$ data set was used to study the reaction $\gamma p \to K^+ \bar K^0 n$ and
 $\gamma p \to p K^+ K^-$, searching for evidence of $S=+1$ resonances in the 
$K^+n$ and $K^+p$ invariant masses.

\begin{figure}
\vspace{6.cm}
\includegraphics{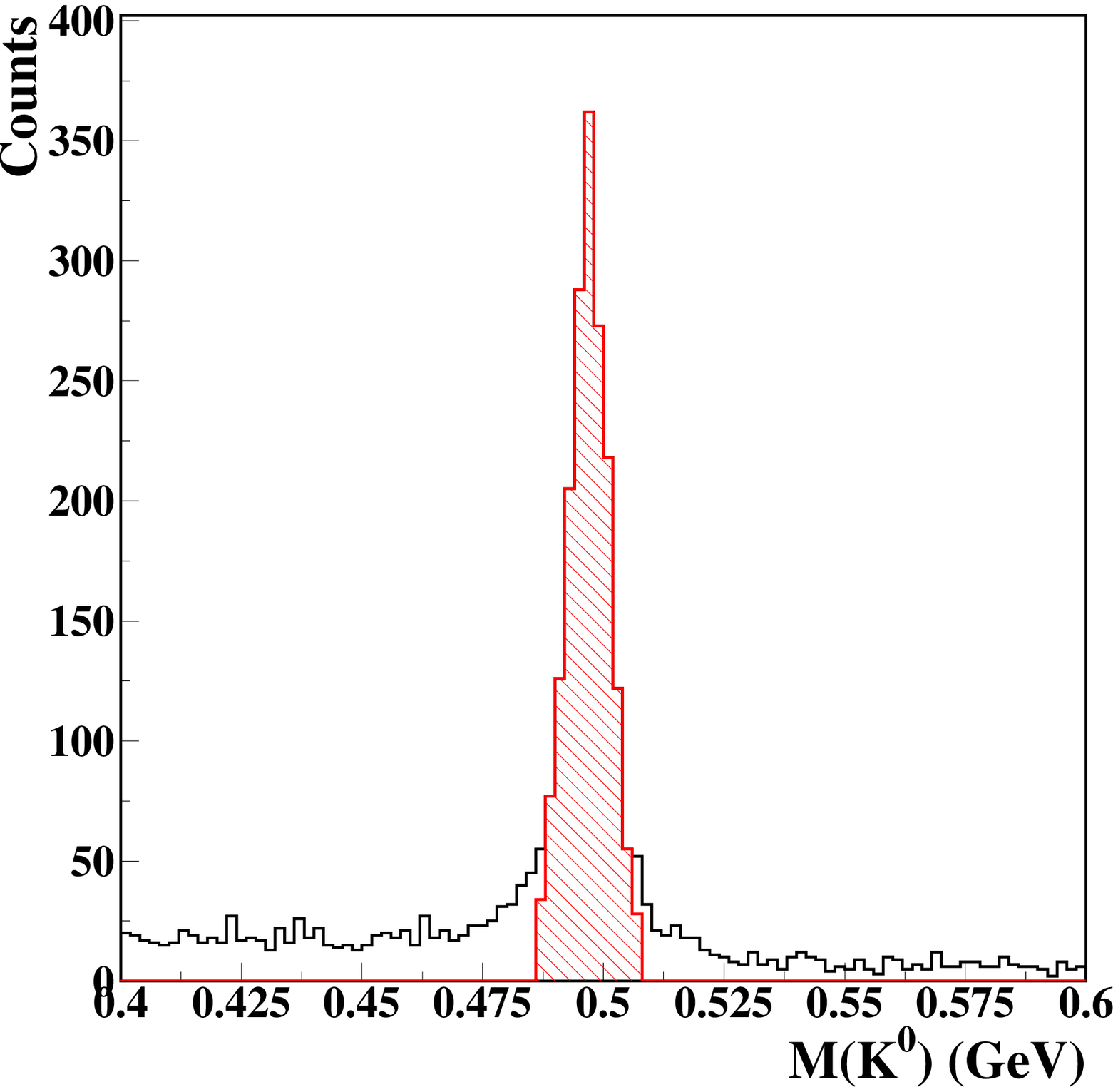}
\includegraphics{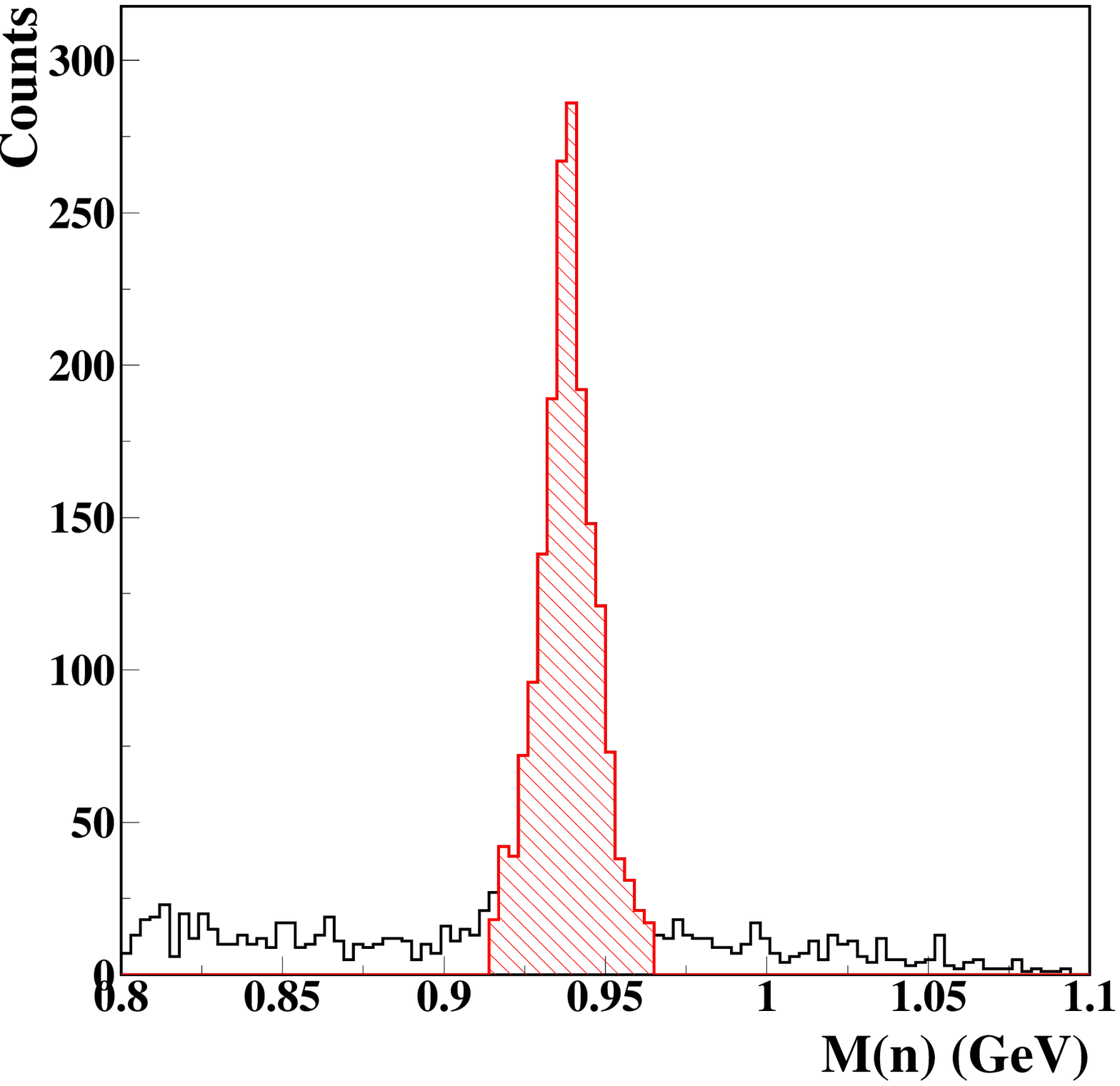}
\caption{Final particle identification for the reaction $\gamma p\to K^+\bar K^0 n$. 
The left plot shows the $\bar K^0$ mass spectrum reconstructed as invariant mass 
the $\pi^+\pi^-$ system. The right plot shows the $K^+\pi^+\pi^-$ missing mass. \label{fig:pid}}
\label{fig:pid_final}
\end{figure}

\begin{figure}
\vspace{6.cm}
\includegraphics{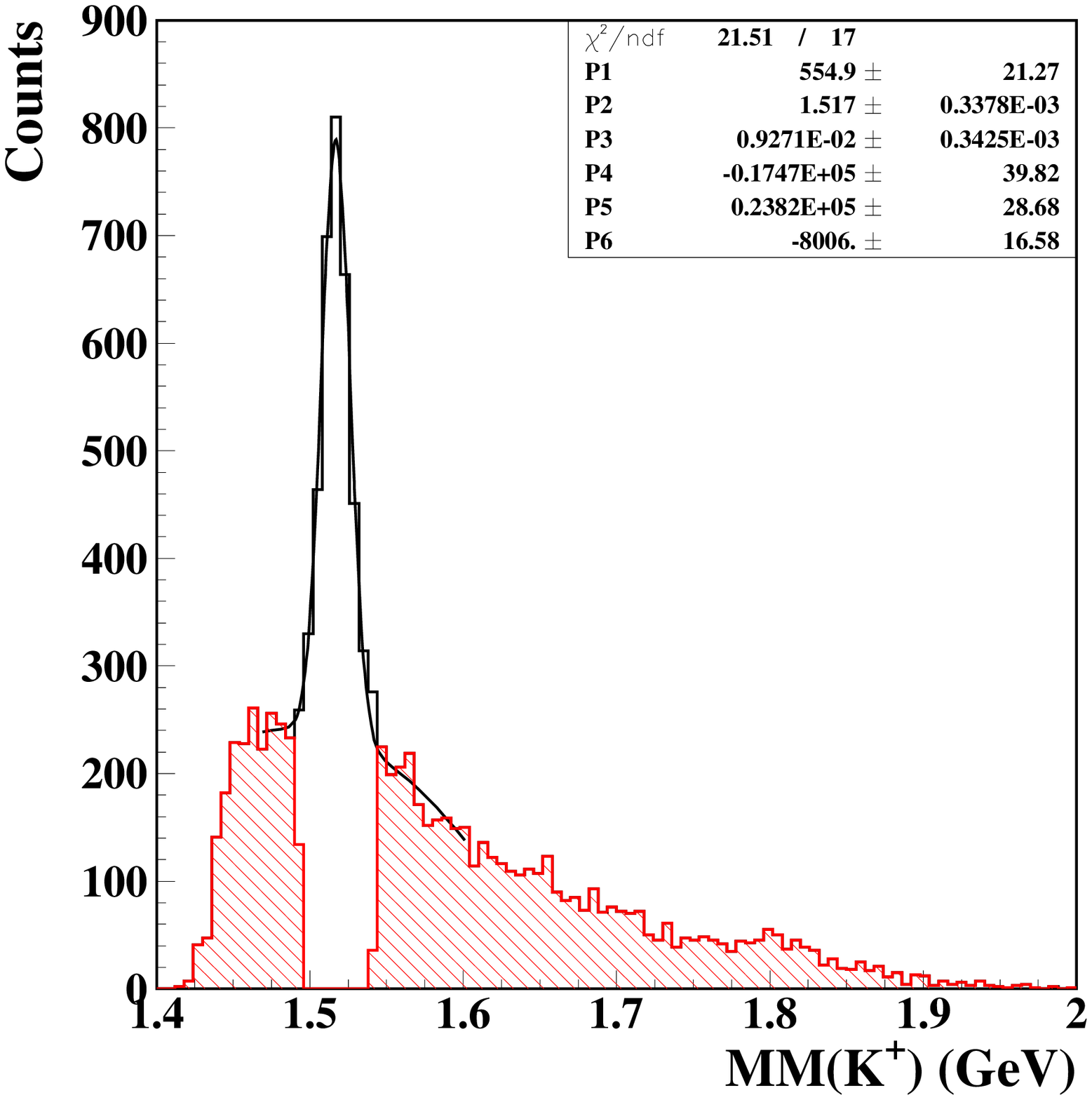}
\includegraphics{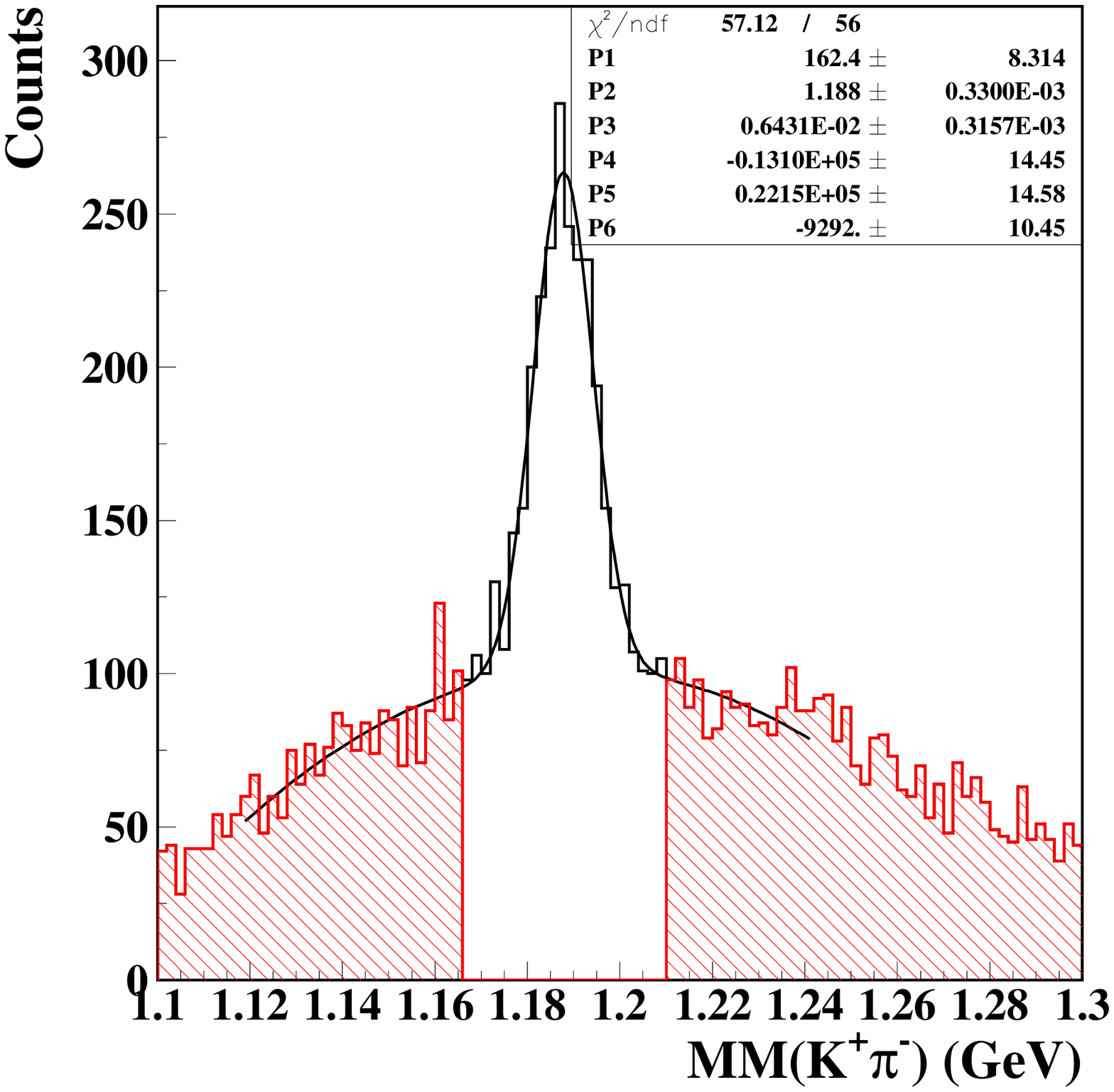}
\caption[]{$K^+$ and $K^+\pi^-$ missing masses after the cuts on the $K^+$
and $\bar{K}^0$, and $n$ masses. The $\Lambda^*(1520)$ and $\Sigma^+$ peaks are clearly visible.
The highlighted  areas correspond to the events selected for further analysis.}
\label{fig:hyperons}
\end{figure}

The first channel was selected by detecting the $K^+$ and reconstructing the 
 $\bar{K^0}$ via the $K_s$ component decaying into $\pi^+\pi^-$. The final state was then 
identified using the missing mass technique. Figure \ref{fig:pid} shows the quality of 
the channel identification: both $\bar K^0$ and $n$ are reconstructed within 1-2 MeV 
of the nominal mass value with small background. The corresponding event sample is 
dominated by the production of known hyperons decaying into the same final state. 
These include $\gamma p \to K^+ \Lambda^*(1520)$, $\gamma p \to K^+ \Sigma^+ \pi^-$, 
and $\gamma p \to K^+ \Sigma^- \pi^+$. Fig. \ref{fig:hyperons} shows the $\Lambda^*(1520)$ 
and $\Sigma^+$ peaks reconstructed as missing mass of $K^+$ and the $K^+\pi^-$ system. Events 
associated with these reactions are excluded by cutting on the corresponding masses. 
After such cuts, the $nK^+$ invariant mass spectrum was constructed. After selecting events 
in which the $\bar K^0$ is emitted at backward angles, two structures with masses  
near $\sim$ 1525 and $\sim 1575$ MeV were visible. However, the low statistics of the 
final event sample did not allow us to draw definitive conclusion on such structures.

\begin{figure}
\vspace{6.cm}
\includegraphics{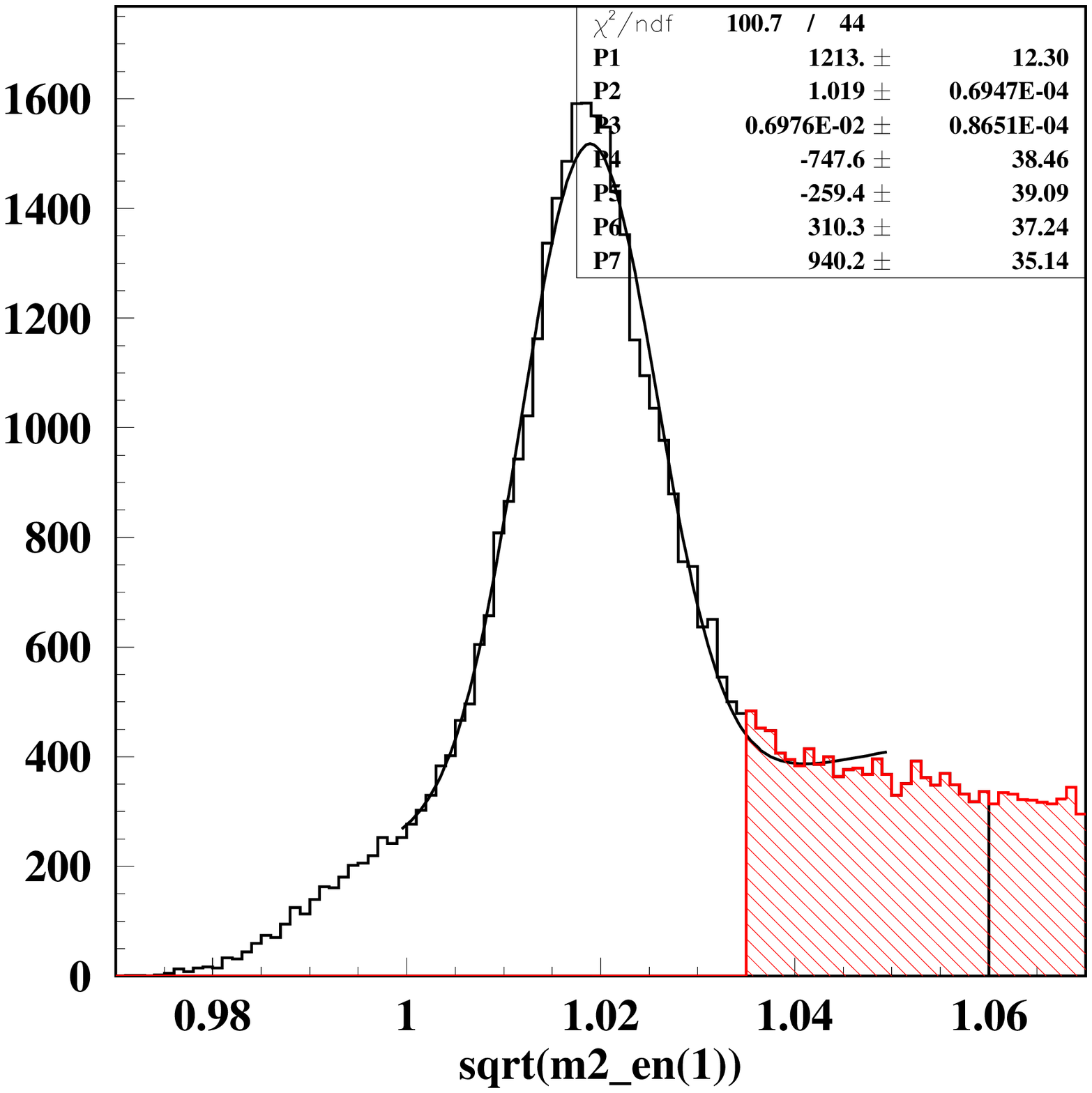}
\includegraphics{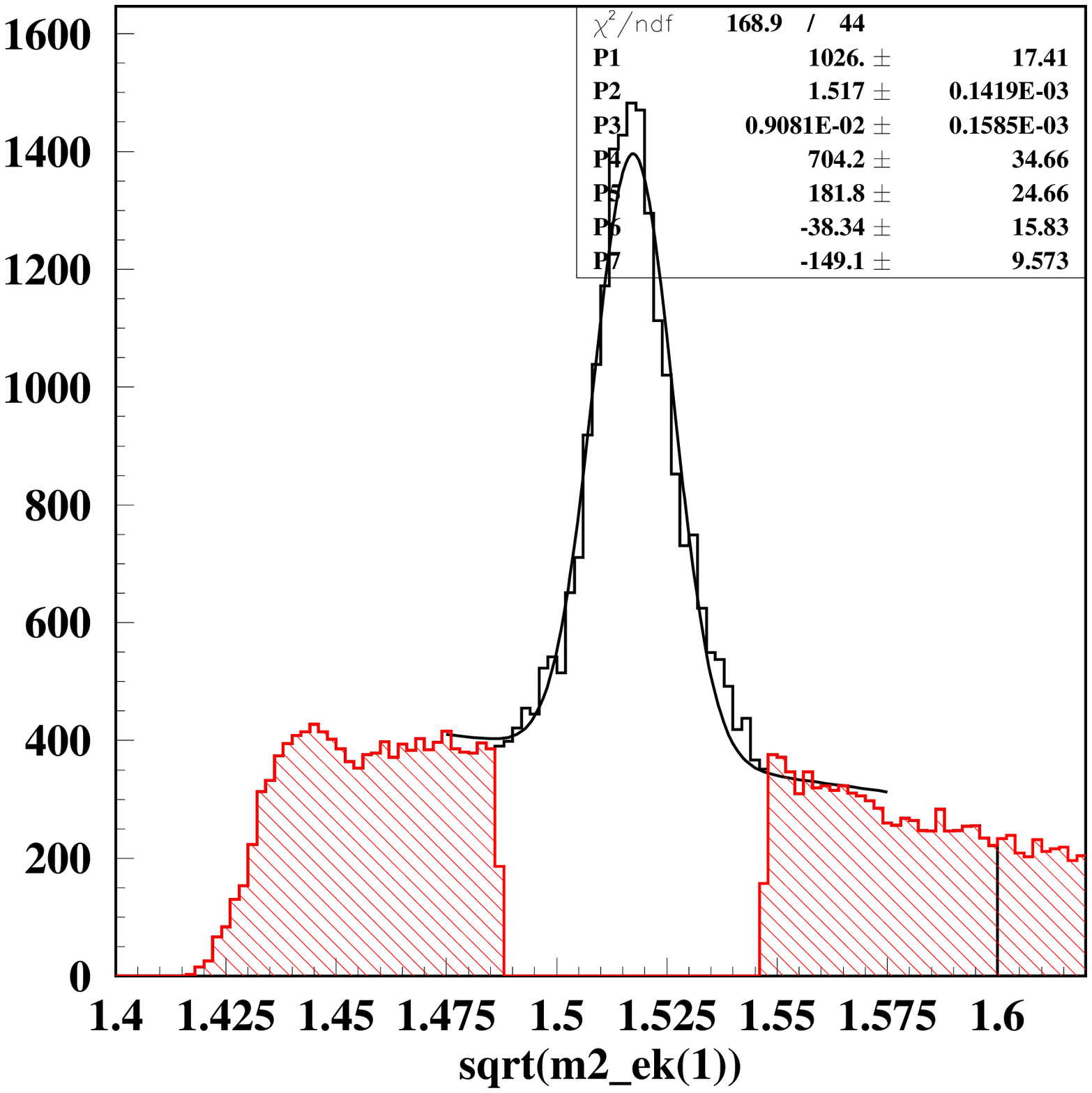}
\caption[]{$p$ and $K^+$ missing masses showing the $\phi(1020)$ and  $\Lambda^*(1520)$
 contribution. The highlighted areas correspond to the events selected for further analysis.}
\label{fig:hyperons2}
\end{figure}

The reaction $\gamma p \to p K^+ K^-$ was selected by detecting two of the three 
charged particles in CLAS and using the missing mass technique allows us to identify the third one. 
Two different topologies, $\gamma p \to p K^+ (K^-)$ and $\gamma p \to (p) K^+ K^-$, were 
analyzed while the topology with $p K^-$ detected was dropped due to the marginal statistics. 
One of the largest contributions to this particular final state is due to the diffractive 
production of the $\phi(1020)$ meson. This is shown in the left panel of Fig. \ref{fig:hyperons2}. 
The associated events as well as the events coming from $\Lambda^*(1520)$ production
 (see the right panel of the same figure) are rejected by cutting on the corresponding masses. 
To further reduce the background contribution coming from other reactions and to maximize 
the signal to background ratio, angular and energy regions were selected where  
Monte Carlo simulations showed maximum sensitivity to the reactions of interest. After 
these additional cuts, the $pK^+$ invariant mass spectrum showed a structure in the 
mass region around 1.58 MeV. However also in this case, the limited statistics did 
not allow us to reach definitive conclusions. 

In either case, the much higher statistics of the g11 experiment that will finish 
data taking by July 29, 2004, will allow more definite conclusions as to the existence 
and significance of these possibly new narrow structures.    

\subsection{Reactions on $^3He$}

\begin{figure}
\epsfxsize=8cm   
\epsfbox{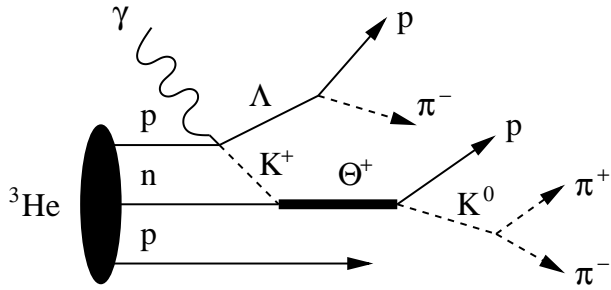}
\caption{ Production mechanism for $\Lambda \Theta^+$ in $^3$He. The decay 
modes $\Lambda \to p \pi^-$, $\Theta^+ \to K^0p$ and $K^0 \to \pi^+\pi^-$ are 
shown here. 
\label{reaction}}
\end{figure}

The $g3c$ data set was analyzed searching for the reaction $\gamma ^3$He$\to p \Lambda \Theta^+$. 
The advantage of this channel (see \cite{guzay} for theoretical predictions 
of the cross section) is that it allows to identify the final state without the need 
of cutting on competing channels, while at the same time excluding kinematical 
reflections in the $NK$ invariant mass spectrum. Moreover, thanks to 
the presence of the $\Lambda$ having strangeness $S=-1$, the $pK^0$ decay mode 
must have $S=+1$.   
The reaction threshold is $E_{\gamma}\simeq 800$ MeV for a $\Theta^+$ mass of 
1.55 GeV/c$^2$. The main reaction mechanism can be pictured as a two-step 
process (Fig.~\ref{reaction}): the initial photon interacts with one of the protons 
of $^3$He and  produces a $\Lambda$ and a $K^+$ ($\gamma p \to K^+\Lambda$). 
The $\Lambda$ leaves the target nucleus, while the $K^+$ reinteracts with the 
neutron in $^3$He to form a $\Theta^+$. In this process, 
one of the two protons of $^3$He can either be a spectator, as pictured in
 Fig.~\ref{reaction}, or rescatter, and thus gain enough momentum to be detected.

The following decay channels are most suitable for detection in CLAS:
\begin{itemize}
\item $\Lambda \to p\pi^-$,
\item $\Theta^+\to pK^0$ and $\Theta^+\to nK^+$,
\item $K^0 \to \pi^+\pi^-$.
\end{itemize}
The final state therefore is $p\pi^- p\pi^+\pi^- p$ for the $\Theta^+\to p K^0$ decay 
mode, and $pp \pi^- nK^+$ for the $\Theta^+ \to nK^+$ decay mode.
Having many particles in the final state (6 for the $\Theta^+\to pK^0$ case, 5 for 
the $\Theta^+\to nK^+$ case), many different topologies of detected particles in 
the final state are possible. 
The most promising three techniques are summarized in Table~\ref{table1}.

\begin{table}[ht]
\tbl{Decay modes, combinations of detected particles in the final state and 
channel-identification techniques adopted for the analysis of the $\gamma ^3$He$\to p \Lambda \Theta^+$ reaction.}
{\begin{tabular}{c  l  l}
\hline
{Decay modes} & {Final-state particles} & {Channel ID}\\
\hline
{$\Theta^+\to pK^0$, $\Lambda\to p\pi^-$} & {$pp \pi^- \pi^+ X $} & {$m_X=m_\Lambda$}\\
{ } & { } & {$m(\pi^-\pi^+) = m_{K^0}$}\\
{ } & { } & { } \\
{ } & {$pp \pi^-\pi^+\pi^- X$} & {$m_X = m_p$}\\
{ } & { } & {$m(p\pi^-) = m_{\Lambda}$ }\\
{ } & { } & {$m(\pi^-\pi^+) = m_{K^0}$}\\
{ } & { } & { } \\
{$\Theta^+\to nK^+$, $\Lambda\to p\pi^-$} & {$pp \pi^- K^+ X$} & {$m_X=m_n$}\\
{ } & { } & {$m(p\pi^-) = m_{\Lambda}$}\\

\hline\hline
\end{tabular}\label{table1}}
\end{table}

\begin{figure}
\begin{center}
\epsfxsize=6.5 cm  
\epsffile{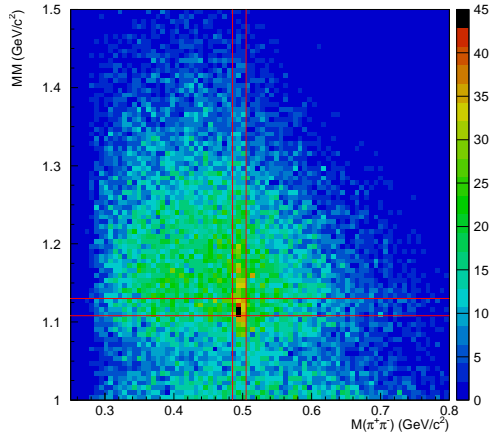}
\caption{Analysis of the $pp\pi^+\pi^-$ topology. The cuts $E_{\gamma}>1$ GeV and 
$p_p<0.8$ GeV/c have been applied in order to reduce the background under the
 $\Lambda$ peak. Shown is here the missing mass of the $pp\pi^+\pi^-$ system as 
a function of the invariant mass of $\pi^+\pi^-$, the lines represent the 
cuts applied to select the $\Lambda$ (horizontal lines) and the $K^0$ 
(vertical lines)\label{miss_lambda}}
\end{center}
\end{figure}

\begin{figure}[thbp]
\vspace{4.cm}
\includegraphics{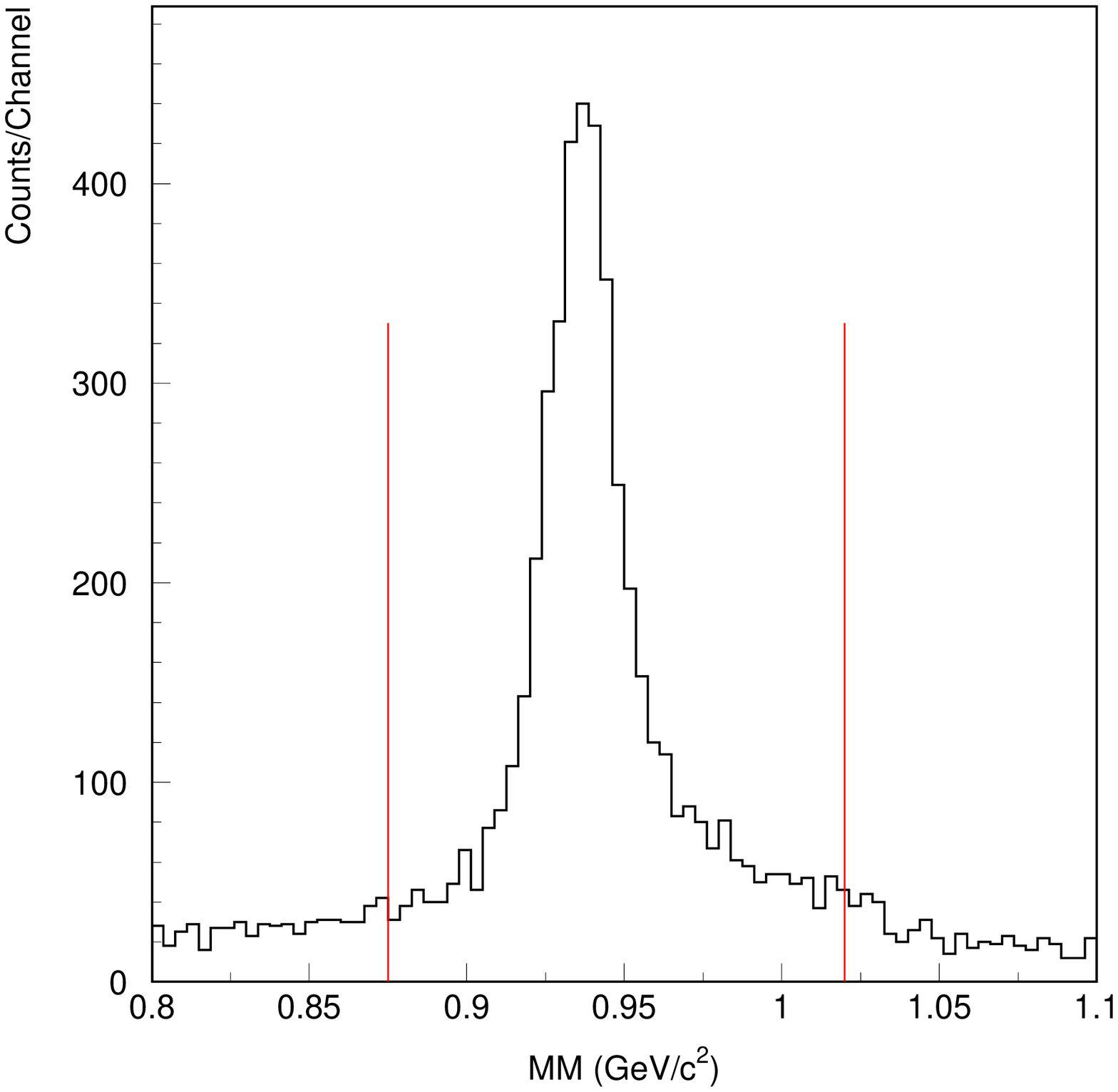}
\includegraphics{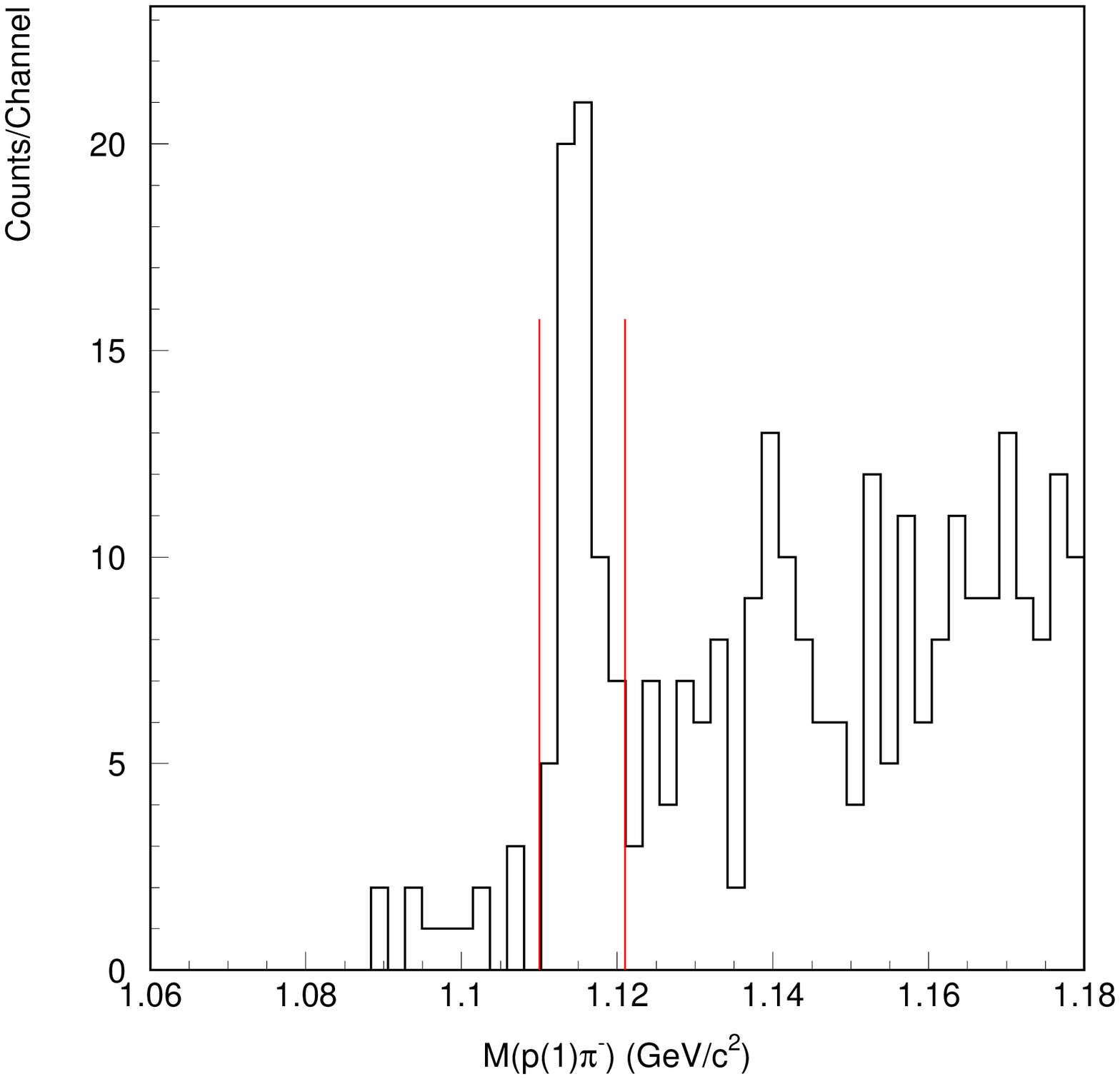}
\includegraphics{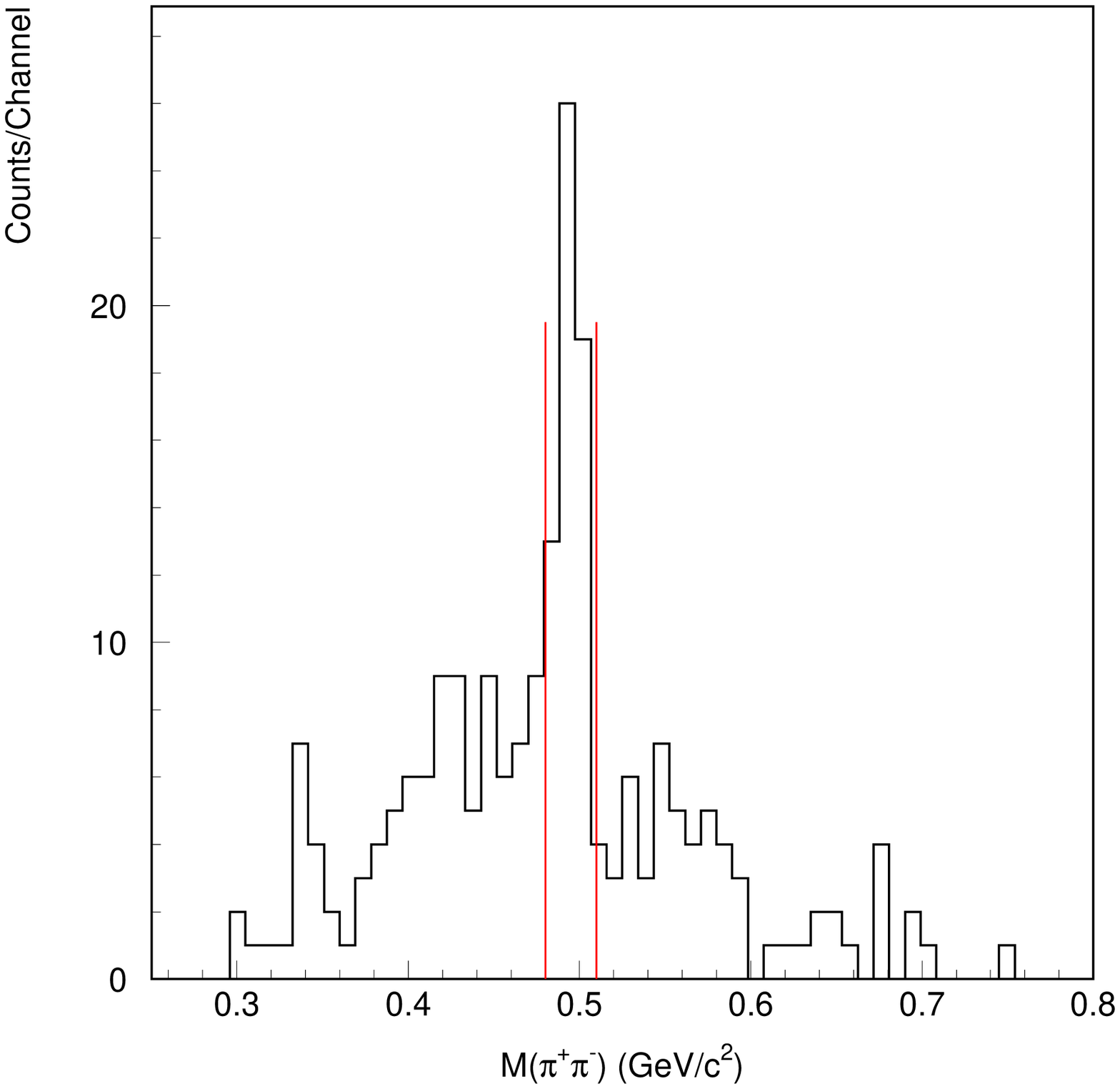}
\caption[]{\small Results of the analysis for the $pp\pi^- \pi^+\pi^-$ topology. Left: missing mass 
of the $pp \pi^- \pi^+ \pi^-$ system, showing a peak a the proton mass; center: invariant 
mass of the $p\pi^-$ system, peaking at the $\Lambda$ mass; right: invariant mass of 
the $\pi^+\pi^-$ system, the peak is at the $K^0$ mass. The vertical lines represent the selection cuts applied.}
\label{2prot}
\end{figure}

\begin{figure}[htbp]
\vspace{4.5cm}
\includegraphics{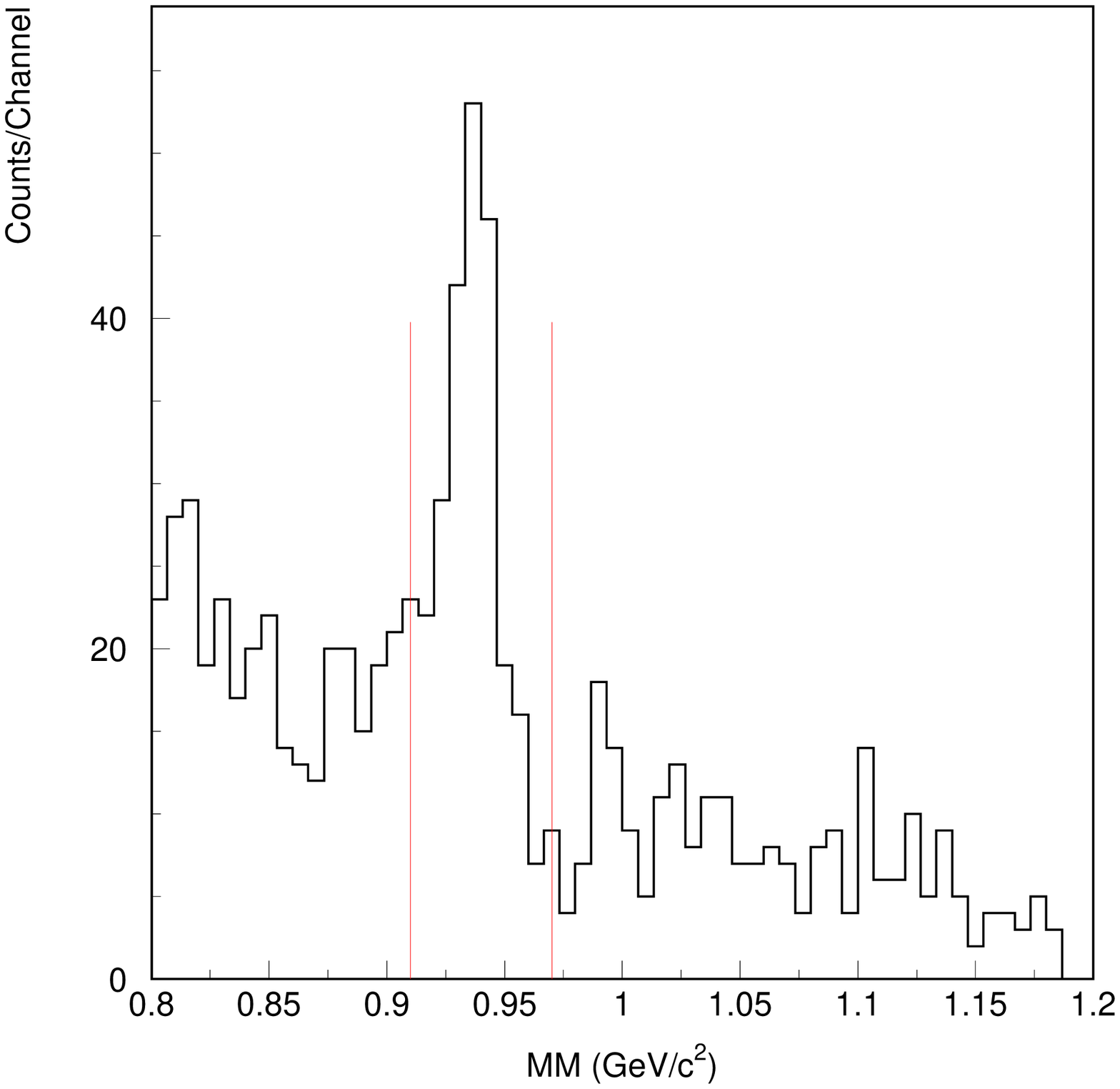}
\includegraphics{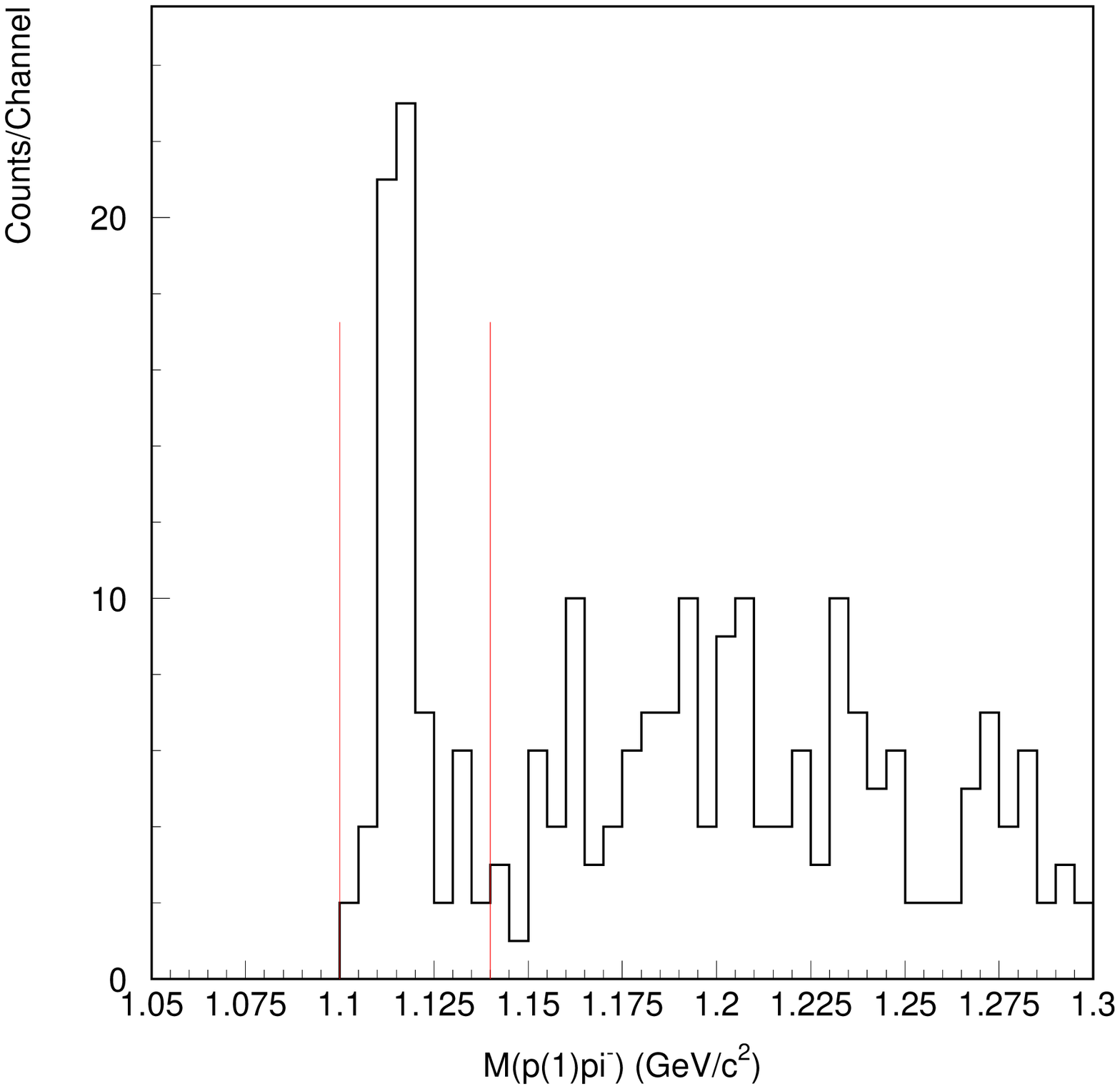}
\caption{\small Results of the analysis for the $pp\pi^-K^+$ topology. Left: missing mass of the 
$p \pi^- K^+$ system, showing a peak at the neutron mass; right: invariant mass of the
 $p \pi^-$ system, peaking at the $\Lambda$ mass. The vertical lines represent the selection 
cuts applied. }
\label{kplus}
\end{figure}

Event selections in the 3 analyses are shown in Figs.~\ref{miss_lambda}, 
\ref{2prot}, \ref{kplus}. In all three cases the final state is well identified, 
without the need of applying cuts to remove background channels. 

The preliminary analysis reveals an enhancement in the $NK$ invariant mass spectrum, near 
1.55 GeV/c$^2$. While the statistical significance of the peak is limited, this 
analysis shows that a search for the $\Lambda \Theta^+$ channel can contribute towards resolving 
the issue of the $\Theta^+$. The final state can be identified, unambigously, and 
no kinematical reflections can produce peaks in the $NK$ invariant mass distribution. An analysis of this 
reaction using the high-statistics photoproduction data on the deuteron (part of the g10 run) 
will give a more definite answer as to the existence and the 
properties of the $\Theta^+$ exotic baryon.

\section{Perspectives \label{perspectives}}

The analysis of existing data shows the capabilities of CLAS to select
 exclusive final states with high multiplicity. The reaction channels described 
above were cleanly identified with small background due to misidentified particles. 
Concurrent reactions decaying to the same final states were seen and rejected from the 
final event sample. However the statistical accuracy of the preliminary data is not 
sufficient to exclude that the observed signals are due to statistical fluctuations, 
kinematic reflections, or some experimental artifact. 
The number of events in the $\Theta^+$ peaks is rather small and 
does not allow us to perform detailed checks of systematic dependencies.

\begin{figure}
\vspace{6.cm}
\includegraphics{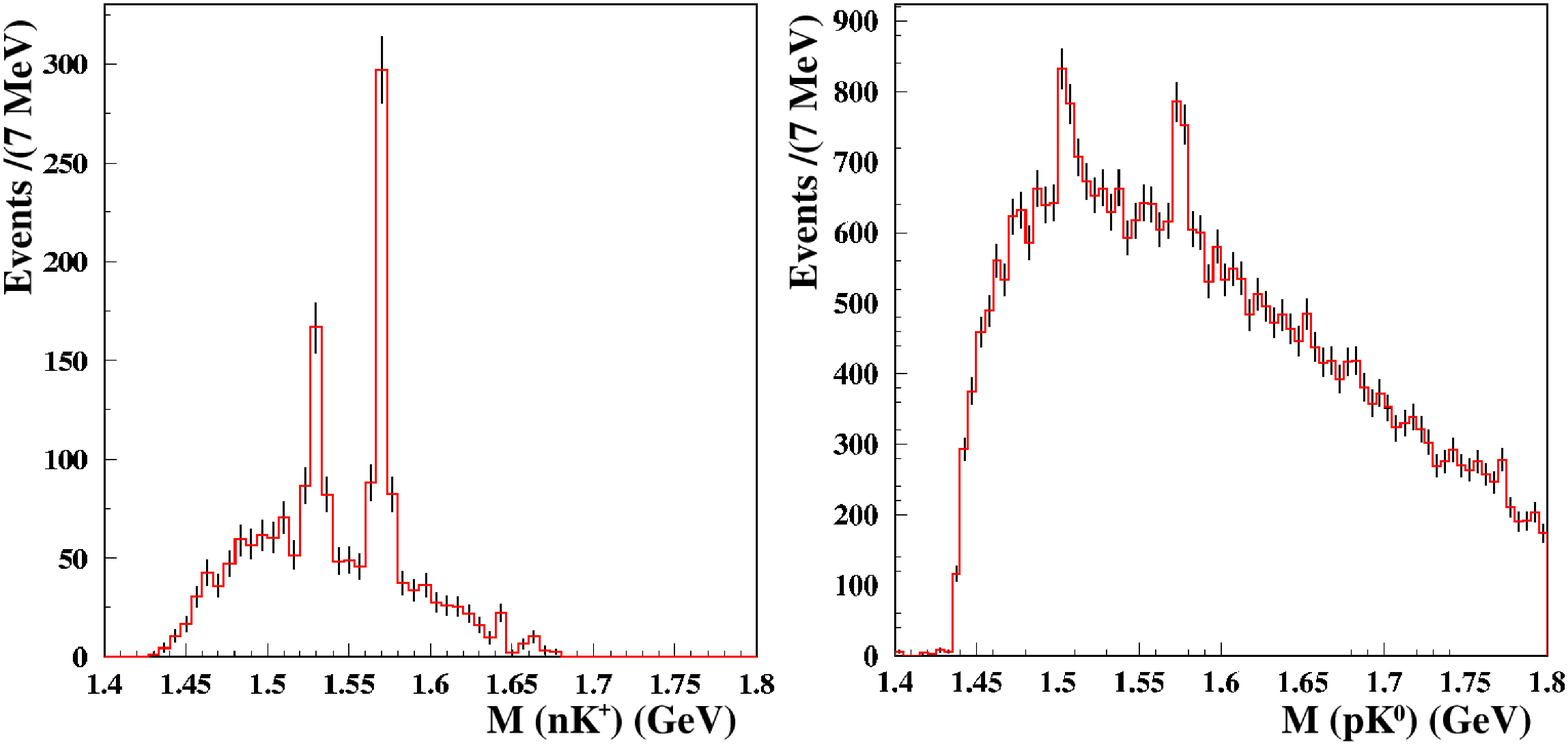}
\caption[]{Expected statistical accuracy of the mass spectra for the reactions 
$\gamma p \to \Theta^+ (\Theta^{+*}) \bar K^0$, with $\Theta^+ (\Theta^{+*})$ 
decaying into $K^+n$ (left) and $pK^0$ (right).}
\label{fig:exp_masses}
\end{figure}

To obtain a definitive answer on the existence of pentaquark states, four dedicated 
experiments were recently approved for CLAS in Hall B at Jefferson Lab. 
The goals and experimental conditions of these experiments are summarized in Table \ref{table:newruns}.

\begin{table}
\tbl{New experiments proposed in Hall B for the search of pentaquark states .}
{\begin{tabular}{|c|c|c|c|l|l|}
\hline
{Run} & {Beam} & {Energy} & {Target} & {Reaction} & {Status}\\ 
\hline\hline
{g10} & {$\gamma$} & {3.8 GeV} & {LD$_2$} & {$\gamma d \to \Theta^+ K^- p$} & {Completed}\\ 
{} & {} & {} & {} & {$\gamma d \to \Theta^+ \Lambda^0$} & {}\\ 
\hline
{g11} & {$\gamma$} & {4.0 GeV} & {LH$_2$} & {$\gamma p \to \Theta^+ \bar K^0$} & {In progress}\\ 
{} & {} & {} & {} & {$\gamma p \to \Theta^+ K^- \pi^+$} & {}\\ 
\hline
{eg3} & {$e$} & {5.7 GeV} & {LH$_2$} & {$\gamma_v p \to \Xi^{--} X$} & {December 2004}\\ 
{} & {} & {} & {} & {$\gamma p \to \Xi^{+} X$} & {}\\ 
\hline
{g12} & {$\gamma$} & {5.7 GeV} & {LH$_2$} & {$\gamma p \to \Theta^+ K^- \pi^+$} & { To be scheduled}\\ 
{} & {} & {} & {} & {$\gamma p \to \Theta^+ \bar K^0$} & {}\\
{} & {} & {} & {} & {$\gamma p \to K^+K^-\Xi^{-}$} & {}\\ 
\hline
\end{tabular}
\label{table:newruns}}
\end{table}

\subsection{Search for the $\Theta^+$ and excited states}

The g10 experiment~\cite{g10}, which has taken data during the spring of 2004, 
aims at studying the production channels $\gamma d \to pK^-\Theta^+$, 
$\gamma d \to p K^0 X$, and $\gamma d \to \Lambda \Theta^+$ with an order of magnitude
improved statistics over the previous g2a run. The g11 experiment will study 
$\gamma p \to \Theta^+ \bar K^0$ and $\gamma p \to \Theta^+ K^- \pi^+$, and 
two decay modes, $\Theta^+ \to nK^+$ and 
$\Theta^+\to pK^0$, increasing by an order of magnitude the statistics of the 
previous data. Both experiments have similar experimental setups and beam 
condition as used in the  g2a and g1c runs, respectively. 
The g11 experiment~\cite{g11} will use tagged photons 
produced from a 4 GeV primary electron beam impinging on a 40-cm long hydrogen target. 
Photons from 1.6 GeV up 3.8 GeV are tagged and the 
data acquisition is triggered by events with at least two tracks to maximize the 
efficiency for the reaction of interest. The total expected integrated luminosity 
is $75$ pb$^{-1}$, i.e. approximately 20 times larger than in the previous run. 

\begin{figure}[hptb]
\vspace{6.cm}
\includegraphics{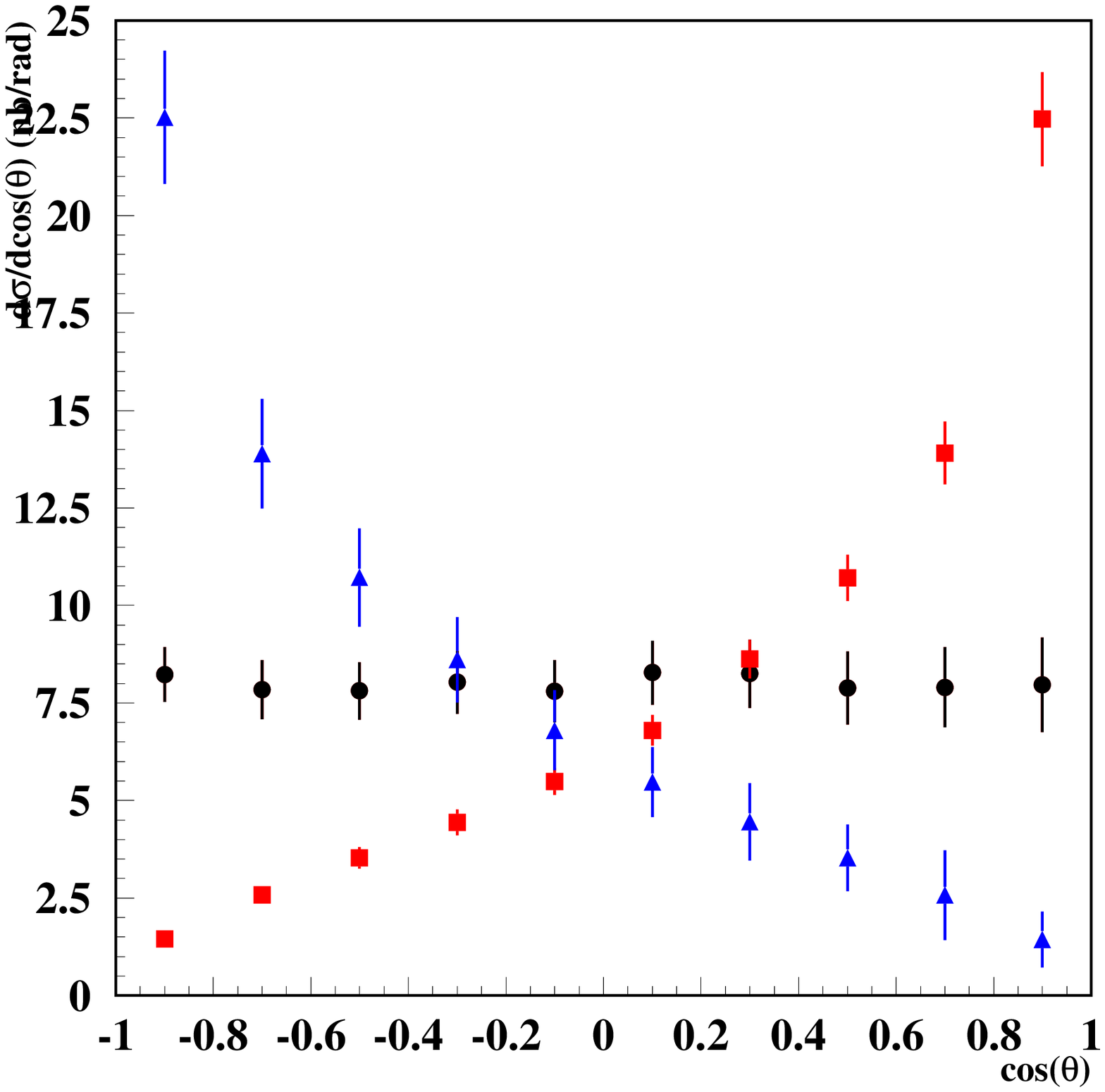} 
\includegraphics{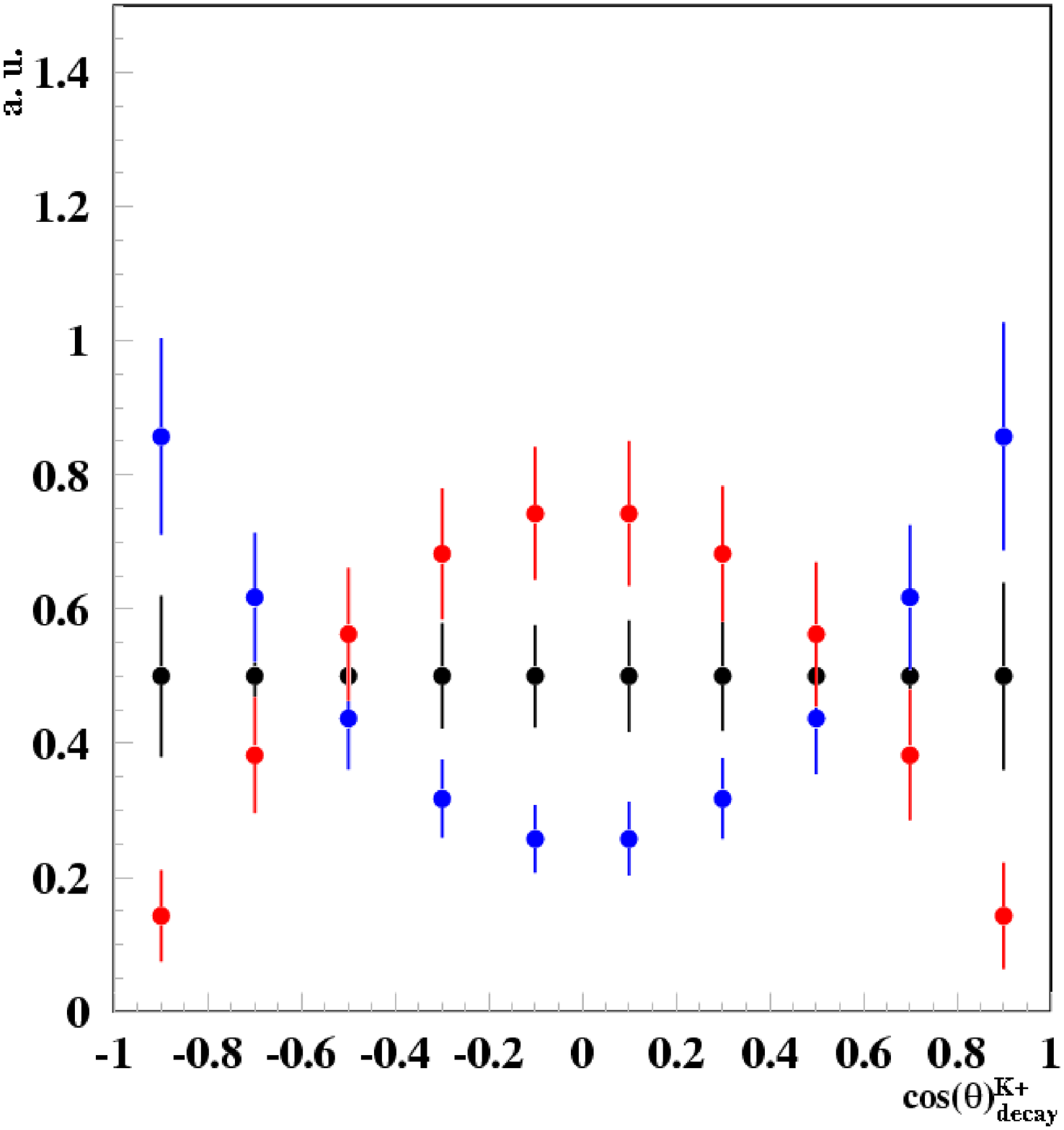}
\caption[]{Expected statistical accuracy for the measurement of the production 
and decay angular distribution. A total cross section of $\sim$ 10 nb was assumed. 
The left plot shows the expected error bars for the production angular distribution 
in the assumption of a t-channel, s-channel, and u-channel production mechanism. 
The right plot shows the expected error bars for the decay angular distribution 
for different assumption on the spin and parity of the state and 
100\% polarization.}\label{fig:exp_xsec}
\end{figure}

If the $\Theta^+$ can be established with certainty, 
the new data will allow us to make progress on establishing the phenomenology of 
the $\Theta^+$ spectrum, e.g. determining in what production channels the 
$\Theta^+$ is seen and what higher mass states are excited. The expected statistical 
accuracy is shown in Fig. \ref{fig:exp_masses}, where the background was estimated 
based on the existing data and the signal was simulated assuming a production cross 
section of $\sim 10$ nb. If the existence of the $\Theta^+$ is confirmed or new 
states are seen, these data will provide accurate measurements of the mass position. In addition, 
the large acceptance of the CLAS detector will allow us to measure both the production 
and decay angular distribution, providing information on the production mechanism and 
spin. Expected statistical accuracy for the measurement of the 
production and decay angular distributions are shown in Fig \ref{fig:exp_xsec}, for 
different assumptions on the production angular distribution and for the spin and 
parity of the pentaquark state.

This measurement will provide a solid foundation for a long term plan for the 
investigation of the pentaquark spectrum and properties.

\subsection{Search for $\Xi_5^{--}$ and $\Xi_5^-$ baryons.} 

The anti-decuplet predicted by the $\chi SM$ or quark-cluster 
models~\cite{jaffe,karliner} for 5-quark states also contains $\Xi_5$ states, 
two of them of exotic nature, the $\Xi_5^{--}$ and the $\Xi_5^{+}$. Evidence for 
such states has so far been seen in only one experiment~\cite{na49}. 
The signal found by NA49 has a mass 100 to 200 MeV away from
 any prediction. This makes it urgent to confirm or refute these claims. Two new 
experiments with CLAS are in preparation to search for  $\Xi_5$ baryons.  

The eg3 experiment~\cite{eg3} will use an untagged electron beam of 5.75 GeV impinging 
on a liquid-deuterium target. Electrons will interact with the neutrons in the 
deuterium through exchange of quasi-real photons. The 
process $\gamma^* n \rightarrow \Xi^{--} X$ will be searched for by 
measuring the decay chain  $\Xi^{--} \rightarrow \pi^- \Xi^- \rightarrow  \pi^- \Lambda \rightarrow \pi^- p$. 
 One proton and 3 $\pi^-$ emerging from three different vertices  have to be reconstructed. 
The experiment is scheduled to take data in the winter 2004/2005.

\begin{figure}[t]
\vspace{6.cm}
\includegraphics{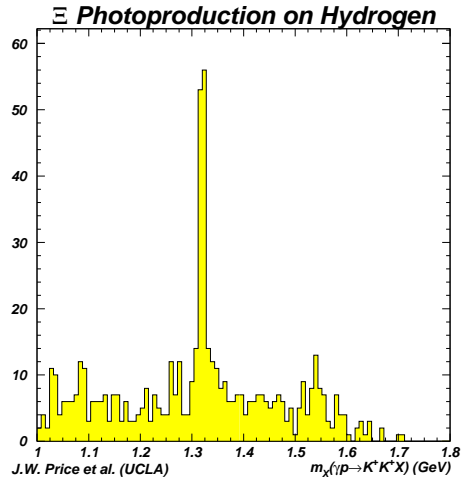} 
\caption[]{Missing mass $M_X$ for the reaction $\gamma p \rightarrow K^+K^+ X$ for 
photon energies in the  range 3.2 to 3.9 GeV. The narrow peak observed for the ground state 
$\Xi(1320)$ illustrates the excellent mass resolution that can be obtained using this method~\cite{price}.
For the pentaquark $\Xi_5$ search, higher energies and much higher statistics are needed. }
\label{fig:cascade}
\end{figure}

The second experiment~\cite{price} is part of g12. It uses the missing mass method 
to search for the  $\Xi^-$ in the exclusive reaction $\gamma p \rightarrow  K^+K^+X$.
If the NA49 results are correct, the $\Xi^-$ would be seen in the missing mass 
spectrum as a peak at 1862 MeV. Excellent missing mass resolution is required for 
such a measurements.
Figure~\ref{fig:cascade} illustrates the method with data taken at photon 
energies between 3.2 - 3.9 GeV. The $\Xi(1320)$ ground state is observed
as a narrow spike. 
Limitations in beam energy and/or in the statistics did not allow us to observe higher 
mass $\Xi's$ in this measurement. 
  
\section{Summary}

In conclusion, CLAS is currently pursuing high statistics searches for the $\Theta^+$ on 
hydrogen and deuterium, and in various final states. We are also searching for possible 
excited states of the $\Theta^+$. The experiments are conducted under similar kinematical conditions as 
previous measurements. The much higher statistics will allow more definite conclusions as to
the existence of the $\Theta^+$ in the exclusive channel $\gamma d \to K^-pK^+(n)$.
In addition, the existence of a possible excited state with mass about 50 MeV above 
the $\Theta^+$ will be clarified. Moreover, an experiment is in preparation to search for 
the $\Xi_5^{--}$ in the mass range where the NA49 experiment at CERN claimed evidence for the 
observation of the $\Xi_5(1862)$ in various charge channels. 
If evidence for the pentaquark states remains and is considerably strengthened, another 
high statistics experiment (g12) will be able to study the spectroscopy of pentaquark states.     

\section{Acknowledgments}
 {\small This work was supported by the Italian Istituto
 Nazionale di Fisica Nucleare, the French Centre National
 de la Research Scientifique, the  French Commissariat
 \`a l'Energie Atomique, the U.S. Department of Energy,
 the U.S. National Science Foundation, and the Korean
 Science and Engineering Foundation. The Southeastern
 Universities Research Association (SURA) operates the
 Thomas Jefferson National Accelerator Facility for the
 United States Department of Energy under contract
 DE-AC05-84ER40150.}

\end{document}